\documentclass[pdflatex,sn-aps]{sn-jnl}

\usepackage{graphicx}%
\usepackage{multirow}%
\usepackage{amsmath,amssymb,amsfonts}%
\usepackage{amsthm}%
\usepackage{mathrsfs}%
\usepackage[title]{appendix}%
\usepackage{xcolor}%
\usepackage{textcomp}%
\usepackage{manyfoot}%
\usepackage{booktabs}%
\usepackage{algorithm}%

\usepackage{algorithmicx}%
\usepackage{algpseudocode}%
\usepackage{listings}%

\usepackage{subcaption}
\usepackage{graphicx}
\usepackage{longtable,booktabs,array}
\usepackage{amsmath}
\usepackage{amssymb}
\usepackage[normalem]{ulem}
\usepackage{bibunits}
\defaultbibliographystyle{sn-aps}
\defaultbibliography{GA_metainterface}
\usepackage[left]{lineno}
\theoremstyle{thmstyleone}%
\theoremstyle{thmstyletwo}%

\theoremstyle{thmstylethree}%

\newcommand{\REV}[1]{{\textcolor{black}{#1}}}

\begin{document}

\title{Automated Discovery of Metainterfaces with Tailored Friction Laws}

\author*[a]{Li Fu}
\equalcont{These authors contributed equally to this work.}
\email{li.fu@ec-lyon.fr}
\author[a]{Djibril Gabriel Kashala}
\equalcont{These authors contributed equally to this work.}
\author[a]{Davy Dalmas}
\author*[a]{Julien Scheibert}
\email{julien.scheibert@cnrs.fr}

\affil[a]{CNRS, Ecole Centrale de Lyon, ENTPE, LTDS, UMR5513, 69130 Ecully, France}

\abstract{Providing dry solid contacts with on-demand macroscale frictional behaviour remains a \REV{formidable challenge} in tribology, haptics or robotics. Metainterfaces created from surfaces with engineered asperity-based topographies can achieve such friction control. However, only few friction behaviours were demonstrated because suitable topographies were identified based on human intuition. Here, we introduce a numerical-optimisation-based inverse design framework to automatically discover new metainterfaces satisfying specified relationships between friction and normal forces (friction law). To illustrate the framework's versatility, we first expand the range of achievable friction coefficients at a constant material pair; we next unlock power-law friction laws with arbitrary exponents between 2/3 and \REV{1.35}; we then achieve bilinear laws with a smaller slope in the second segment than in the first. We validate relevant cases experimentally. By enabling systematic exploration of large parameter spaces, not limited to topography but potentially incorporating the individual asperities' bulk material or surface physicochemistry, our automated framework offers design solutions for any physically possible friction law. It also provides new insights into the elusive relationship between local interfacial properties and macroscopic friction.
}

\maketitle




Friction is an ambivalent force that humanity has always sought to master. Sufficiently high friction is necessary for locomotion or manipulation~\cite{costa2022} while minimizing friction is key to reduce the 20\% of the world's total energy production that is consumed to overcome friction~\cite{holmberg2017}. The frictional resistance of an interface is often summarized by the friction coefficient, i.e., the slope of the (supposedly proportional) evolution of the friction force, $F$, vs the confining normal force, $P$. However, many current technologies require increasingly sophisticated frictional behaviour from functional contacts, for instance in robotics to securely grasp various objects~\cite{bicchi2000,liu2023d}, or in haptic displays and virtual reality to convincingly emulate the tactile perception of diverse surface textures~\cite{millet2017, basdogan2020}. In such cases, achieving specified, non-linear evolutions of the entire function $F(P)$, so-called friction law, is desirable.

Unfortunately, the interplay of mechanical and physicochemical processes that determine friction along a given contact interface is so complex and intricate~\cite{vakis2018} that the optimisation of functional contacts remains mostly done through long and costly trial-and-error processes to identify adequate surface materials and textures. Despite undeniable individual successes of that empirical approach \cite{scheibert2009, murarash2011, baum2014, li2016, huang2025}, we still lack a generic framework that would deliver a suitable interface design for any set of physically possible specifications on the friction law. 

A first step in this direction was recently made on the class of elastic contacts, with the introduction of so-called metainterfaces~\cite{aymard2024}. The surface texture of those interfaces can be engineered to offer specified, non-linear friction laws. However, so far, metainterface designs were identified for very few types of friction laws, where inverse design could be carried out analytically, based on human intuition. While providing valuable proofs of concept, such approach has left most of the design space of metainterfaces unexplored. 
6

In this context, developing a generic framework for automated discovery of surface designs that shape friction laws would trigger a crucial acceleration in how solid contact interfaces are conceived and functionalized. Those design solutions could be further used to build a deeper understanding of the \REV{elusive} relationship between surface micro-texture and macroscale friction. To tackle those fundamental and applied challenges, we formulate the hypothesis that automated surface design can be achieved by combining the metainterface approach with multi-objective optimisation methods. Here, we test this hypothesis on three different sets of frictional specifications. We also experimentally validate some of the proposed new designs. 

\section*{Design Strategy}\label{design-strategy-and-friction-model}

\subsection*{Metainterfaces and friction model}
\begin{figure}[H]
     \centering
     \includegraphics[width=\linewidth]{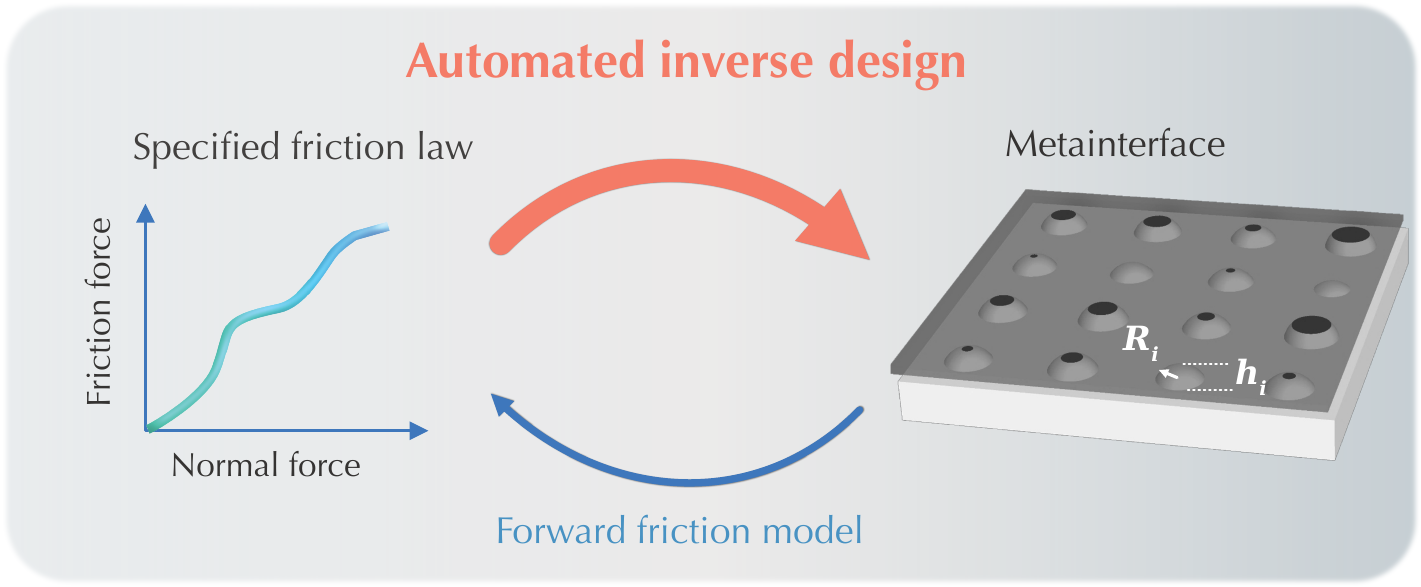}
        \caption{Metainterfaces offer on-demand friction behaviours thanks to their engineered asperity-based surface geometry (right). The lists of asperity heights $h_{i\in[1,N]}$ and curvature radii $R_{i\in[1,N]}$ determine contact (black disks) and friction through a tribology-informed forward model. Here, the model is automatically inverted using numerical optimisation, providing designs meeting a specified friction law (left).}
\label{fig-illustration}
\end{figure}

Metainterfaces is a concept that was recently proposed~\cite{aymard2024}, which enables quantitative control of an interface's friction law through the design of its surface topographies (Fig.~\ref{fig-illustration}). In such metainterfaces, one surface is composed of $N$ asperities, each with a well-defined individual geometry, while the counter-surface is smooth. The design parameters are the geometrical descriptors of all asperities (shape and height with respect to the base plane). The impacts of the materials, physicochemical surface phenomena and small-scale roughness on friction are accounted for by a preliminary experimental calibration of the compression and shear behaviours of individual microcontacts. The macroscale friction response is then modeled as the collective response of the population of $N$ microcontacts. This multi-asperity friction model is finally inverted to identify surface designs that offer a desired friction law.

In~\cite{aymard2024}, this generic design framework was applied to the contact between a rigid plane and a designable array of independent spherical elastic asperities. These shared a common material, characterized by a composite elastic modulus $E^*$ ($E^*$=$\frac{E}{1-\nu^2}$, with $E$ and $\nu$ being Young's modulus and Poisson's ratio of the elastic material) and a common radius of curvature $R$.  The design quantity to be determined was the list of individual heights, $h_i$, of the $N$ asperities. Applying Hertz's theory \cite{barber2018}, the contact area under pure compression of the $i^{\text{th}}$ microcontact, $a_{0,i}$, is a function of the separation, $\delta$, between the rigid indenting plane and the base plane of the asperities:
\begin{equation}
  a_{0,i} = 
\pi R (h_i - \delta) \quad \text{if } h_i \geq \delta, \qquad a_{0,i} =0 \quad \text{if } h_i < \delta.
\label{Eq:area}
\end{equation}
In~\cite{aymard2024}, the macroscale friction force, $F$, was found proportional to the contact area under pure compression, $A_{\text{0}}$=$\sum^N_{i=1} a_{0,i}$, through a coefficient $\sigma B$ that was found constant ($\sigma$ is the friction stress of the interface and $B$=$A_{\text{f}}/A_{\text{0}}$, where $A_{\text{f}}$ is the (reduced) contact area when sliding occurs). In this system, the normal and friction forces, $P$ and $F$, are described by the following set of analytical equations~\cite{aymard2024}:
\begin{equation}
P = \sum_{i=1}^{N} \frac{4E^*}{3\pi^{3/2}R} a_{0,i}^{3/2} ;\qquad
F =  \sigma A_{\text{f}} = \sigma B A_{\text{0}} = \sigma B \sum_{i=1}^{N} a_{0,i}.\label{Eq:F}
\end{equation}
In this article, $P$, $F$, $A_{\text{0}}$ and $A_{\text{f}}$ are made dimensionless by defining: $\tilde{P}$=$3P/(4 N E^* R^{1/2} h_{\text{m}}^{3/2})$, $\tilde{F}$=$F/(N \pi \sigma B R h_{\text{m}})$, $\tilde{A}_{\text{0}}$=$A_{\text{0}}/(N \pi R h_{\text{m}})$ and $\tilde{A}_{\text{f}}$=$A_{\text{f}}/(N \pi B R h_{\text{m}})$, with [0, $h_{\text{m}}$] the range of possible asperity heights. Such a finite range is inevitable in experimental interfaces, due to the finite number of asperities and the maximum manufacturable height. By construction, $\tilde{F}=\tilde{A}_{\text{0}}=\tilde{A}_{\text{f}}$ (Eq.~\ref{Eq:F}) and $\tilde{h}=h_i/h_{\text{m}}\in[0,1]$.

\subsection*{Optimisation-based inverse design}

The purpose of the inverse design step is to start from a desired list of constraints on the $F(P)$ relationship, and to invert Eqs.~\ref{Eq:area}-\ref{Eq:F} to find the $N$ values of the individual $h_i$ that satisfy the constraints, thus providing the desired friction law.

Navigating the high-dimensional, non-intuitive design space of frictional metainterfaces requires using computational optimisation algorithms. The field of computational materials design offers a diverse suite of optimisation techniques, each with distinct strengths and weaknesses \cite{cerniauskas2024,zhai2024}. Gradient-based methods \cite{aage2017,bordiga2024} offer rapid convergence but are limited to differentiable objective functions, \REV{making them ill-suited for our problem involving contact mechanics discontinuities (contact/non-contact transitions).}
Generative models~\cite{bastek2023, nordhagen2025, mouton2026}, which learn a direct mapping from property to structure, require the significant upfront cost of generating a massive training database \REV{and making them provide solutions that strictly satisfy a target physical law beyond statistical resemblance is notoriously difficult and computationally expensive~\cite{zampini2025}.}

\REV{In this context, we have considered alternative classes of solution-search algorithms and selected the Genetic Algorithm (GA, see Methods for details on its implementation) as a particularly well-suited method for our metainterface inverse design problem, based on a benchmark study between four methods (three metaheuristics and one probabilistic, see SI section 1). GA is} a population-based search algorithm employing operators such as selection, crossover, and mutation to iteratively evolve a population of candidate solutions toward an optimal state \cite{katoch2021}. While GAs have already been used in tribological context~\cite{cinat2020}, their suitability for the specific challenge of metainterfaces stems from several key attributes. First, as a global optimisation method, a GA maintains a diverse population of solutions, making it robust against premature convergence to local optima. Second, for GAs, the fitness of a candidate solution (quantified score of the match between current and target friction laws) can be evaluated without requiring the function to be continuous or differentiable. Third, GAs are inherently parallelizable and highly effective at exploring large, complex design spaces involving either discrete or continuous variables. 

\subsection*{Experimental validation}
While GA-based automated inverse design provides a high throughput of new designs, their experimental realization (Methods) is comparatively costly, so we have selected only some of them for experimental comparison. For each selected design, a polydimethylsiloxane (PDMS) elastomer sample bearing $N=64$ asperities with the designed heights ($h_{\text{m}}$ is set to 120$\mu$m) is produced. \REV{The sample dimensions and pitch of the square array of asperities satisfy the recommendations made in~\cite{zeka2026} for the assumption of independent hertzian contacts to be valid (SI section 5).} Compression and friction tests are performed  against a rigid glass plate (a kind of interface shown in~\cite{aymard2024} to satisfy all the assumptions underlying Eqs.~\ref{Eq:area}-\ref{Eq:F}). The geometrical ($R$), material ($E^*$) and interfacial parameters ($B$ and $\sigma$) entering Eqs.~\ref{Eq:area}-\ref{Eq:F} are calibrated as described in Methods. \REV{Typically twenty} tests at varying normal loads $P$ enable measurements of the laws $A_{\text{0}}(P)$, $A_{\text{f}}(P)$ and $F(P)$, where static and dynamic friction are distinguished (Methods). \REV{SI sections 7 and 8 show that the measured behaviour laws are negligibly affected by manufacturing errors and wear, respectively.}

\section*{Proofs of concept}\label{results}

We present the design solutions obtained when targeting three types of friction laws, each illustrating different capabilities of our automated framework.

\subsection*{Case 1: Proportional friction law - New insights on the friction coefficient}\label{case-study-1-linear-friction-law}

We target metainterfaces that exhibit proportional friction laws, i.e., $F$=$\mu P$ with the slope $\mu$ of the law defining the friction coefficient. Such a proportional law is a cornerstone in tribology, and the friction coefficient, the most widely used quantifier of friction, is often considered as a characteristic of any given material pair. In contrast, here we target various friction coefficients at constant material pair. We note however that a perfect proportionality between $F$ and $P$ is fundamentally impossible at very small $P$ for asperity-based contacts. Indeed, the first contact always occurs through a single contact with the highest asperity, giving rise to non-linear evolutions of $A_{\text{0}}$ and $F$ with $P$, according to the non-linear Hertz model ($A_{\text{0}}$$\sim$$P^{2/3}$). Thus, one can only expect a proportional law in an asymptotic sense, when $P$ is sufficiently large to involve enough microcontacts.

In this context, we apply our GA optimisation framework to the fitted linear trend ($\tilde{F}$=$a\tilde{P}+b$) of the final 20\% points of the dimensionless friction law $\tilde{F}(\tilde{P})$. The final point of the law is defined at $P_{\text{max}}$, the necessary $P$ to bring the lowest asperity into contact. We perform the optimisation to target two objectives: an arbitrary asymptotic slope $a$ between 1 and 4, and a vanishing intercept $b$, using fitness functions detailed in Methods.

\begin{figure}[H]
     \centering
     \begin{subfigure}{\textwidth}
         \centering
         \includegraphics[width=\textwidth]{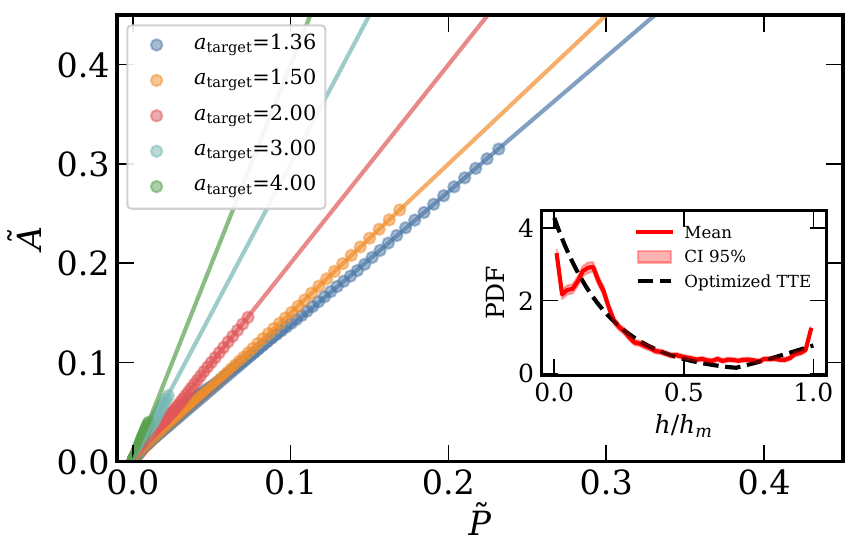}
     \end{subfigure}
     \hfill
     \begin{subfigure}{\textwidth}
         \centering
         \includegraphics[width=\textwidth]{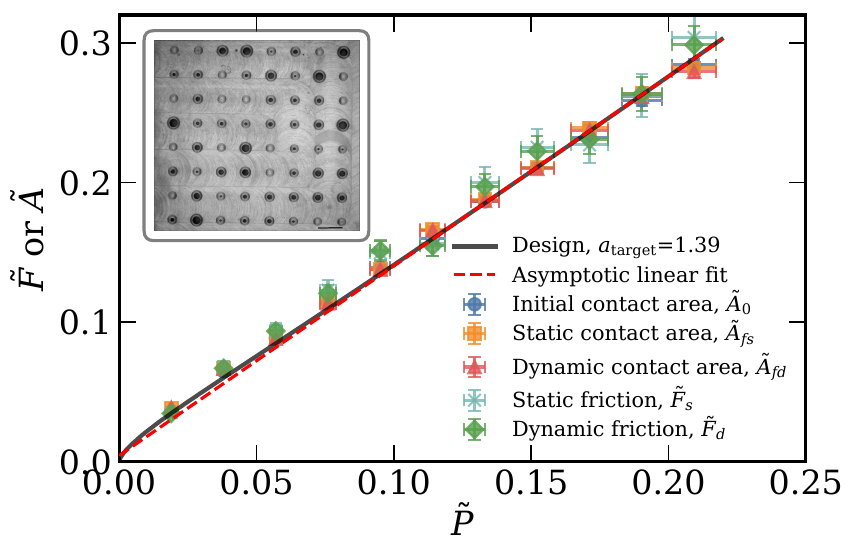}
     \end{subfigure}
        \caption{Metainterfaces with proportional friction laws. (A) Main: examples of designed friction laws (disks) for target slopes from 1.36 to 4. Linear fits of the curves (solid lines) show that the fitted $a$ always matches $a_{\text{target}}$ to better than 0.1\%, and that the fitted intercept $b$ is always smaller than 0.1\% $a_{\text{target}}P_{\text{max}}$, demonstrating good proportionality. Inset: \REV{Solid red line (resp. red band) is the average (resp. 95\% confidence interval) of the PDF of the designed asperity heights over 400 independent optimisation runs for $a_{\text{target}}$=1.39}. Black dashed line: optimised continuous TTE distribution for the same $a_{\text{target}}$.
        (B) Experimental validation on all measurable quantities (see legend and Methods) for $a_{\text{target}}$=1.39. Error bars: SD of 10000 different calculations where each quantity entering $\tilde{A}$, $\tilde{P}$ or $\tilde{F}$ is drawn from Gaussian distributions with mean (resp. SD) equal to the values (resp. error bars) provided in SI section 4. Black solid line: designed behaviour (red dashed line: linear fit of the final 20\% points). Linear fits of all data points over $\tilde{P}$ $\in$ [0.07-0.22] show that the experimental slope matches that of the designed law (fitted over the same range) to better than 3.8\%. Inset: image of the metainterface under $\tilde{P}$=0.15. \REV{Scale bar: 1.5\,mm.}
        }
\label{fig-linear_friction_law}
\end{figure}

Figure~\ref{fig-linear_friction_law}A shows examples of the obtained solutions, demonstrating designs giving nicely proportional laws with a large range of slopes. As shown in Fig.~S9, a slope accuracy better than 0.1\% can be achieved for all $a_{\text{target}}$ $>$1.12. For the same range of slopes, the intercept is also minimal (less than 0.1\% of $a_{\text{target}}\tilde{P}_{\text{max}}$), i.e. we achieve a good asymptotic proportionality and not just a linear trend.
Below 1.12, the non-linearity at small $P$ is too strong to be compensated by any list of asperity heights. Good proportionality over the whole range $[0,P_{\text{max}}]$, quantified by a goodness of proportional fit $R^2$ $>$0.99 is obtained for all target slopes above $a_{\text{min}}$=1.36. This limit sets the minimum targetable friction coefficient to $\mu_{\text{min}}$=$\frac{3 \pi \sigma B \sqrt{R}}{4 E^* \sqrt{h_{\text{m}}}}a_{\text{min}}$.

The inset of Fig.~\ref{fig-linear_friction_law}A represents the \REV{average} probability distribution function (PDF) of the asperity heights over \REV{400} design solutions for the same target slope but different initializations of the GA. The PDF is very stable across runs (narrow red band in Fig.~\ref{fig-linear_friction_law}A, inset), indicating a robust underlying average solution. Analyzing such solutions for all tested slopes, we could propose an analytical description of the PDF made of two contributions (Methods): a decaying exponential at small $h$, and a triangular tail at large $h$ that mitigates the initial non-linearity of the friction law. Such a triangular-tailed exponential (TTE) PDF of heights involves only three dimensionless parameters (standard deviation, SD, of the exponential ; width of the triangle ; areal weight of the triangle in the full PDF), rather than the $N$ parameters of the height list. With this simplified description, the design space becomes much smaller, making GA-based optimisations one order of magnitude faster. 
The design of quasi-linear friction laws shown in~\cite{aymard2024} being a particular case of TTE distribution (vanishing triangular tail), applying GA on this single-parameter problem (the exponential's SD) shows that proportional laws of the same quality as for the full TTE (slope accuracy$<$0.1\%, intercept $<$0.1\%$a_{\text{target}}\tilde{P}_{\text{max}}$, global $R^2$$>$0.99) are possible only for $a_\text{target}$$>$2.25.
TTE distributions, with $a_\text{min}$=1.36, thus unlock a larger range of friction coefficients than existing designs \REV{(see SI section 3)}.

Figure~\ref{fig-linear_friction_law}B shows the experimentally measured behaviour of one of those 3-parameters-based design solutions (list of heights in Tab.~S3). We have selected a slope of 1.39, close to the minimum accessible slope (1.36, i.e. $\mu_{\text{min}}$ $\simeq$1.00 for the experimental parameters of Fig.~\ref{fig-linear_friction_law}B) while keeping a good proportionality, both to challenge our automated design, and to enable quantitative comparison with one previous design from~\cite{aymard2024}. As expected, the dimensionless measured laws $\tilde{A}_{\text{0}}(\tilde{P})$, $\tilde{A}_{\text{f}}(\tilde{P})$ and $\tilde{F}(\tilde{P})$ overlap within the error bars. Most importantly, they do match quantitatively the designed  behaviour (black line). The fitted value of the slope of the measurements closely matches that of the design (caption of Fig.~\ref{fig-linear_friction_law}B), while the intercept remains very small. For this case of $a_{\text{target}}$=1.39, the intercept for the friction law designed in~\cite{aymard2024} was about 4 times larger than that achieved with the present TTE-based design, demonstrating its superiority over previous, empirical designs.

\subsection*{Case 2: Generalized nonlinear power friction law}\label{case-study-2-non-linear-power-law-friction-response}

While proportional friction laws are often assumed in engineering design, various natural and technological systems exhibit non-linear frictional behaviours, in particular power laws where $\tilde{F}$$\sim$$\tilde{P}^{k}$, with $k$ $\neq$1~\cite{fusco2004, toannguyen2014, liu2023c}. A trivial example of power-law metainterface is when all asperities have the same height, thus producing a power friction law with the Hertz exponent of 2/3 (experimentally, it is the case of our calibration interfaces, Fig.~S12A). In this context, the ability to design  metainterfaces offering power friction laws, but with any prescribed value of the exponent $k$ \REV{in a range including} 2/3 (Hertz-like law) and 1 (proportional law), would provide a generalisation of the known friction laws. 

\begin{figure}[H]
     \centering
     \begin{subfigure}{1\textwidth}
         \centering
         \includegraphics[width=\textwidth]{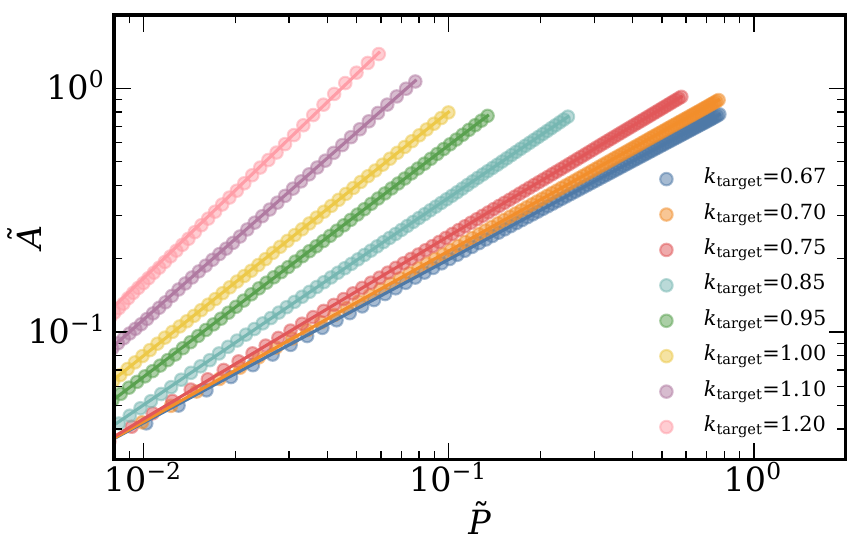}
     \end{subfigure}
     \hfill
     \begin{subfigure}{1\textwidth}
         \centering
         \includegraphics[width=\textwidth]{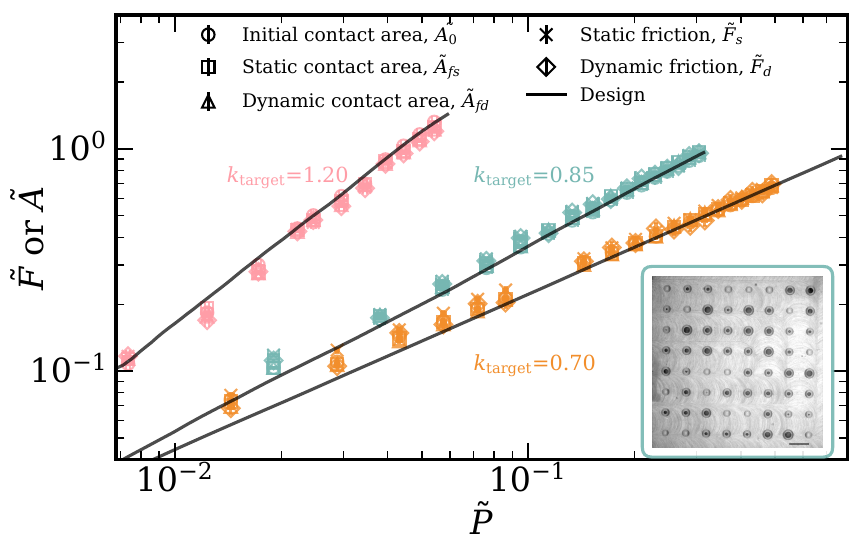}
     \end{subfigure}
     \caption{\label{fig-power_friction_law}Metainterfaces with power friction laws. (A) Designed friction laws (disks) and power law fits (solid lines) for various target exponents (legend). The linear trends in log-log scales show power law behaviours. Fitted $k$ are always matching $k_{\text{target}}$ to better than 0.7\%. (B) Experimental validation for $k_{\text{target}}$=\REV{0.70, 0.85 and 1.20}. Plotted quantities and error bars are as in Fig.~\ref{fig-linear_friction_law}B. Black solid lines: designed behaviours. Power law fits of all data points show that the exponents match those of the designed laws to better than 7.7, 4.5 and 3.4\%, respectively. Inset: image of the metainterface \REV{with $k_{\text{target}}$=0.85} under $\tilde{P}$=0.15. \REV{Scale bar: 1.5\,mm.} In both panels, the curves are vertically shifted by a factor of $100^{(k_\text{target}-2/3)}$ for readability.
       }
\end{figure}

To achieve such generalisation, we apply our automated design and perform a two-objectives optimisation (Methods) combining a target exponent, $k_{\text{target}}$, and a measure of the goodness of the power-law fit of $\tilde{F}(\tilde{P})$. We consider $k_{\text{target}}$ between \REV{0.50} and \REV{1.50}. Figure~\ref{fig-power_friction_law}A shows that those targets can be successfully achieved, \REV{with an exponent error less than 0.1\% and a coefficient of determination, $R^2$, larger than 0.99, provided that $k_{\text{target}} \in$[0.67 ; 1.35] (Fig.~S10). Below 2/3, no solution is found, suggesting that it represents a physical lower limit imposed by the underlying Hertz-based contact model. Above, solutions exist but with a degraded quality.}

To validate those solutions, the experimentally measured laws $\tilde{A}_{\text{0}}(\tilde{P})$, $\tilde{A}_{\text{f}}(\tilde{P})$ and $\tilde{F}(\tilde{P})$ are shown in Fig.~\ref{fig-power_friction_law}B for \REV{metainterfaces with target exponents spanning the accessible range: 0.70, 0.85 and 1.20 (lists of heights in Tabs.~S4--6)}. Quantitative agreement is found with the designed laws (solid lines) within the error bars, and the fitted target exponents closely match that of the design (caption of Fig.~\ref{fig-power_friction_law}B). This agreement underscores the capacity of the numerical optimisation method to discover non-intuitive surface topographies for achieving specific functional targets. The asperity height distributions that produce \REV{those} specific non-linear power \REV{responses were} found complex (Fig.~S5, right) and dependent on $k_{\text{target}}$. While an analytical representation and interpretation of such complex height distributions is beyond the scope of this paper, it is a compelling avenue for future investigation.

\subsection*{Case 3: Bilinear friction laws - Designing both the height and radius of each asperity}\label{case-study-3-bi-linear-friction-laws-with-a-functional-transition-point}

In this last case, we target segmented friction laws, \textit{i.e.}, providing different friction behaviours for different ranges of normal force. In~\cite{aymard2024}, bilinear friction laws were obtained but, unfortunately, they only covered the case where the second linear segment had a larger slope than the first one. Here, we overcome this limitation and obtain bilinear friction laws with a potentially smaller slope in the second segment.

\begin{figure}[h!]
     \centering
     \begin{subfigure}[b]{0.9\textwidth}
         \centering
         \includegraphics[width=\textwidth]{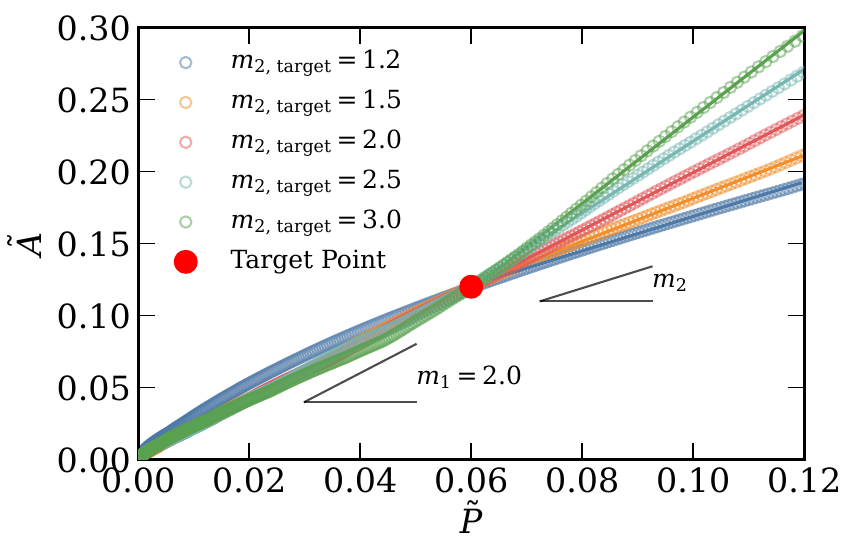}
     \end{subfigure}
     \hfill
     \begin{subfigure}{0.95\textwidth}
         \centering
         \includegraphics[width=\textwidth]{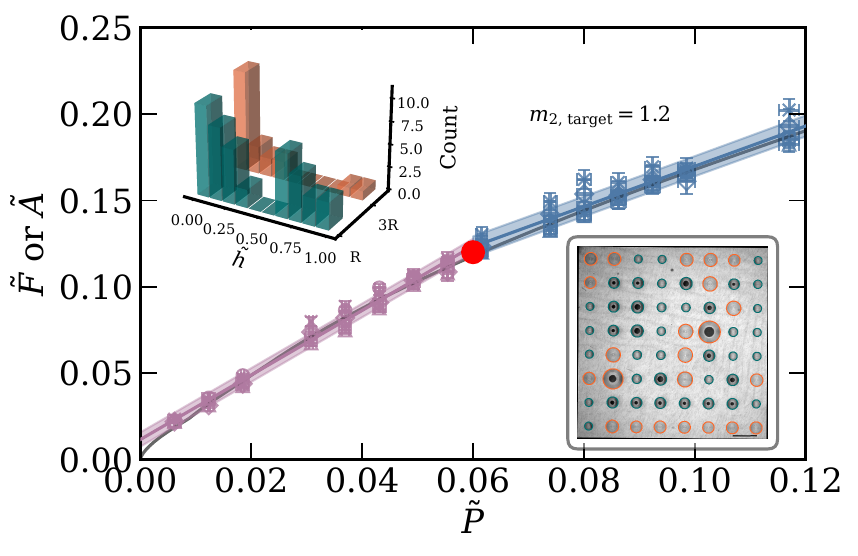}
     \end{subfigure}
        \caption{\label{fig-branches_friction_law} Metainterfaces with bilinear friction laws. Large red disk: target operating point through which the law must pass (here imposing an initial slope $m_1$=2.0). (A) Typical designed friction laws (disks) and linear fits (solid lines) of the second segments ($\tilde{P}\in [0.06, 0.12]$) for various design target slopes $m_\text{2,target} \in [1.2-3.0]$ (legend). $m_\text{2,target}$ is always obtained to better than 1\% and can be either larger or smaller than $m_1$. \REV{(B) Experimental validation for $m_{\text{2,target}}$=1.2. Plotted quantities and error bars are as in Figs.~\ref{fig-power_friction_law}B. Black solid line: designed behaviour. Colored lines (resp. colored bands): linear fits  (resp. 95\% prediction interval) of the data points on both sides of the target point (large red disk). The fitted slope of the second segment matches that of the designed law to better than 2.5\%. Lower inset: image of the metainterface under $\tilde{P}$=0.062. Asperities with radius $R$ (resp. 3$R$) are circled in green (resp. orange). Scale bar: 1.5\,mm. Upper inset: distribution of heights and radii (same color code as in lower inset).}}
\end{figure}

To achieve this, we increase the size of the design space by optimising not only the height, $h_i$, but also the radius of curvature, $R_i$, of each asperity, with \REV{two} possible values of the radius \REV{($R$ and 3$R$)} (Methods). Such a doubling of the number of design parameters has no impact on how the GA optimisation works. The design solutions are however more complex, because now consisting of a coupled probability distribution of both $h$ and $R$ (see e.g. upper inset of Fig.~\ref{fig-branches_friction_law}B).

Figure~\ref{fig-branches_friction_law}\REV{A} demonstrates that bilinear friction laws are indeed obtained, and that the slope of the second segment can not only be larger, but also smaller than that of the first one. This opens new possibilities for applications that require
sophisticated, force-dependent friction control.

\REV{Figure~\ref{fig-branches_friction_law}B shows an experimental validation of the design with the smallest target slope for the second segment in Fig.~\ref{fig-branches_friction_law}A (1.2, list of heights in Tab.~S7). Despite the additional challenge of manufacturing and calibrating asperities with two different curvature radii (Methods), the agreement between measured and designed laws is quantitatively as good as for the previous cases using a single radius (Figs.~\ref{fig-linear_friction_law}B and~\ref{fig-power_friction_law}B).}

\section*{Discussion}\label{discussion}

Our automated inverse design framework has successfully demonstrated its ability to discover metainterfaces with precisely tailored friction laws, on cases defying human intuition \REV{(see SI section 3 for a representation of the design space unlocked by our work on Cases 1 and 2)}. Relevant cases have been tested experimentally and show quantitative agreement with the prescribed tribological behaviour, on both contact area and friction. 

Beyond offering black-box solutions, our numerical inverse designs provide valuable physical insights by generating numerous different surface topographies that all equally satisfy a given friction law (inset of Fig.~\ref{fig-linear_friction_law}A, Fig.~S5 and SI section 2.3). Analyzing this ensemble reveals whether the solutions are statistically diverse or share systematic height‑distribution features essential for performance. Our investigation of proportional laws has, e.g., identified a non-intuitive effect: a group of asperities with a large height increases the unavoidable low‑load nonlinearities, but also enables a later inflection that drives the curve toward the desired asymptotic proportionality. This unforeseen solution is qualitatively distinct from the one obtained empirically in~\cite{aymard2024}, which did offer desired slopes, but with an irreducible intercept, making the asymptotic friction law only linear, not proportional.

Collectively, those results validate our initial hypothesis that combining the metainterface concept with a numerical optimisation of the asperities' geometry can significantly advance the quest of quantitative friction control, as long as the inverted friction model is carefully fed by experimental calibrations at the microcontact level.

\REV{The friction model used here is arguably the simplest model that captures the calibrated behaviour of our specific PDMS/glass interfaces, in the limited range of contact and friction conditions representative of our target friction laws (including the specific steady driving velocity and range of normal force). Addressing more complex target behaviours involving, e.g., sliding transients, would necessarily require more advanced calibrations like those typically used to identify the parameters of rate-and-state friction laws~\cite{baumberger2006}. Also, using different materials, scales or manufacturing methods would likely modify the dominant interfacial phenomena; viscoelasticity, adhesion, finite strains or elastic interactions may become relevant and require additional, dedicated calibrations. The friction model that will comprehend all those calibrated behaviours will presumably become more complex, possibly irreducible to simple analytical expressions like those of Eqs.~\ref{Eq:area}-\ref{Eq:F}, thus requiring sophisticated physics-based models \cite{papangelo2019}, finite element simulations \cite{lengiewicz2020,zeka2026}, or experimental lookup tables.}

\REV{While such foreseen complexity would have been prohibitive in the context of the intuition-based analytical inverse designs performed in~\cite{aymard2024}, the present fully numerical optimisation approach represents an ideal versatile framework to handle virtually any form of calibrated behaviour, friction model, target law and objectives. Although GA can be used with any of those forms, other optimisation methods may become more efficient and a new benchmark study would be useful to identify the most relevant search algorithm.}

Overall, our automated design framework represents an easy-to-implement, virtually universal tool for designers of elastic, friction-based functional devices, that is expected to be useful in fields including sport equipment, biomedical devices, haptics or soft robotics.

\newpage
\section*{Methods}
\subsection*{GA implementation}
An individual within the GA population is represented by a chromosome, i.e., a vector of $N$\REV{=64} real-valued genes, $[\tilde{h_1}, ..., \tilde{h_N}]$, where $\tilde{h_i}$ is the dimensionless height of asperity $i$.
An initial population (500 individuals) is generated by sampling each gene from a uniform random distribution over $[0,1]$. The fitness of each individual is evaluated based on the desired friction law. The evolutionary process proceeds through iterative generations, producing offspring via selection, crossover, and mutation operators. We use the Non-dominated Sorting Genetic Algorithm II (NSGA-II) \cite{deb2002}, implemented using the DEAP library~\cite{fortin2012}. The NSGA-II selection mechanism uses non-dominated sorting and crowding distance metrics to rank individuals, promoting convergence toward the Pareto-optimal front while maintaining population diversity. Offsprings are generated using two-point crossover (crossover probability 0.4) and polynomial bounded mutation (mutation probability 0.9; mutation distribution index 0.5). 
\REV{Such high mutation rate enhances exploration of the rugged fitness landscape over the parameter space. The optimisation terminates when the maximum fitness value of each objective (see below) in the population stabilizes (improvement $<10^{-5}$ over 50 generations) or a maximum of 1000 generations is reached.}
\REV{
All optimisations are repeated for at least 30 independent runs. The convergence stability, reproducibility, scalability, trade-offs and solution quality of the GA-based optimisations are discussed in SI section 2.
}

\textbf{Fitness functions for proportional friction laws} - To simultaneously maximize the slope accuracy and minimize the intercept of the fitted asymptotic friction law ($\tilde{F}$=$a\tilde{P}+b$), the fitness functions are 
$f_1$=$1/(1 + |a - a_{\text{target}}|/a_{\text{target}})$ and  $f_2$=$1/(1 + |b|)$. $f_2$ is activated only when the slope mismatch falls below 1\%, ensuring the primary focus is to obtain the correct slope. \textbf{Fitness functions for power friction laws} - To maximize both the exponent accuracy and the $R^2$ goodness of the nonlinear least-squares regression ($\tilde{F}$=$m \tilde{P}^k$) of $\tilde{F}(\tilde{P})$ over the whole range of $P$, the fitness functions are $f_1$=$1/(1 + |k - k_{\text{target}}|/k_{\text{target}})$ and $f_2$=$1/(1 + (1-R^2))$. $f_2$ is activated only when the exponent mismatch falls below 1\%, ensuring the primary focus is to obtain the correct exponent. \textbf{Fitness functions for bilinear friction laws} - The curve $\tilde{F}(\tilde{P})$ is segmented at the target point $[\tilde{P}_{\text{target}}, \tilde{F}_{\text{target}}]$=$[0.06, 0.12]$. Each segment is linearly fitted. The first segment slope $m_1$ is $\tilde{F}_{\text{target}}/\tilde{P}_{\text{target}}$=2. To simultaneously minimize (i) the minimum distance $d_{\text{min}}$ between the curve and the target point and (ii) the mismatch with the target slope $m_{2,\text{target}}$ in the second segment, and maximize the goodness of fit in both segments ($R^2_1$ and $R^2_2$), the fitness functions are $f_1$=$1/(1 + 10d_{\text{min}})$ ; $f_2$=$1/(1 + 10(1-R^2_1))$ ; $f_3$=$1/(1 + 10|m_2 - m_{2,\text{target}}|)$ ; $f_4$=$1/(1 + 10(1-R^2_2))$.

\subsection*{Experimental methods}
\textbf{Sample preparation} - Parallelepipedic blocks (thickness 7.2\,mm, lateral sizes 20$\times$20\,mm) of PDMS (Sylgard 184), attached to a glass plate, are crosslinked following the protocol recommended in~\cite{delplanque2022} (24\,h at 25$^{\circ}$C, demolding, and 24\,h at 50$^{\circ}$C). The flat surface, decorated by a square lattice (pitch 1.5\,mm) of 8×8=64 spherical asperities, replicates an aluminum mold featuring spherical holes drilled with a 500\,$\mu$m radius sphere-ended tool. \REV{The standard deviation of the manufacturing error on each asperity's height is $\sim$2.1\,$\mu$m (64 measurements, see SI section 7).} The counter-surface is a clean, smooth glass plate.
\textbf{Mechanical testing} - The experiments are performed in a clean room (temperature 23$\pm$1$^{\circ}$C, humidity 36$\pm$4\%) using the setup described in Fig. 13 of \cite{guibert2021}. Forces are acquired at 195\,Hz, with resolutions of $\pm$7\,mN for $P$ and $\pm$(1\,mN+0.01$Q$) for the tangential force $Q$. Images of the real contact (9.3 to 10.4\,$\mu$m/pixel) are taken as in \cite{sahli2018}. The contact area of each microcontact is measured by simple grey-level thresholding (threshold defined individually using Otsu’s method \cite{otsu1979} applied to the image at maximum normal force).
\textbf{Calibration of individual microcontacts} -
\REV{For each of the two PDMS batches, and for each of the two nominal asperity radii,} single microcontact calibration is performed on a sample featuring five identical asperities of height 170\,$\mu$m, distant by more than 7.1\,mm. Individual contact areas and forces are denoted $a_{\text{0}}$, $a_{\text{f}}$, $p$ and $f$, while their macroscopic counterparts in metainterfaces are $A_{\text{0}}$, $A_{\text{f}}$, $P$ and $F$. Uncertainty on $a$ is $\pm$0.0028$\sqrt{\pi a}$\,mm$^2$. Compression tests (steps of amplitude 1\,$\mu$m and duration 30\,s) providing $a_{\text{0}}(p)$ are fitted with Hertz's theory (Fig.~S12A) to provide the value of $E^*$, knowing the curvature radius $R$ measured from profilometry measurements on the five asperities. \REV{The good agreement with Hertz indicates that, in the present experiments, adhesion is negligible, presumably due to the milling-process-induced residual roughness of the asperities, $1.5\pm0.4\,\mu$m rms (64 measurements on the sample used in Fig.~\ref{fig-linear_friction_law}B), a value sufficient to drastically reduce pull-off forces in elastomer contacts~\cite{fuller1975}). } Based on shearing tests, the shear-induced area-reduction ratii at static and dynamic friction (see below), $B_s$ and $B_d$ respectively, are fitted on $a_{\text{f,s}}(a_{\text{0}})$ and $a_{\text{f,d}}(a_{\text{0}})$, using a proportional law (Fig.~S12B). Asperity-level proportionality between $f$ and $a$ is demonstrated in Fig.~S12C. \REV{The calibrated values of $R$, $E^*$, $B_s$ and $B_d$ used in Figs.~\ref{fig-linear_friction_law}B, \ref{fig-power_friction_law}B and \ref{fig-branches_friction_law}B are given in Tab.~S1.}
\textbf{Macroscale behaviour laws} - After aligning the PDMS and glass surfaces, shearing tests are performed under constant $P$ over a distance of 1\,mm, at constant velocity 0.1\,mm/s (images taken at 10\,fps). For each $P$, static ($F_{\text{s}}$) and dynamic friction ($F_{\text{d}}$) are respectively measured as the peak of $Q$ versus tangential displacement and the average of $Q$ over the displacement range $[0.5$$-$$0.8]$\,mm (in full sliding). The contact areas $A_{\text{f,s}}$ and $A_{\text{f,d}}$ are measured similarly, in addition to the initial contact $A_{\text{0}}$. For each metainterface, $\sigma_s$ and $\sigma_d$ are determined from the proportional fits of $F_{\text{s}}(A_{\text{f,s}})$ and $F_{\text{d}}(A_{\text{f,d}})$ under normal forces $P$$>$$P_{\text{max}}$ (\REV{see values in Tab.~S2}). The indices s and d are omitted in all general statements throughout the article, except in Figs.~\ref{fig-linear_friction_law}B, \ref{fig-power_friction_law}B and \ref{fig-branches_friction_law}B.

\subsection*{Analytical friction law with TTE height distribution}

We consider the following asperity height distribution, $\phi(h)$:
\begin{equation}
\phi(h)=\left\{
    \begin{array}{ll}
    \frac{\gamma}{\lambda} e^{-\frac{h}{\lambda}} & \text{if} \, h\in[0,h_{\text{m}}-b],\\
    \frac{\gamma}{\lambda} e^{-\frac{h}{\lambda}} +\frac{2 S_{p}}{b^2} (h-h_{\text{m}}+b) & \text{if} \, h\in[h_{\text{m}}-b,h_{\text{m}}],\\
    \end{array}
    \right.\label{eq:distrib}
\end{equation}
\noindent and $\phi(h)$=0 elsewhere. $b$ is the width of the triangular peak in the distribution, which has its maximum at $h_{\text{m}}$ and vanishes at $h_{\text{m}}$-$b$. $S_{p}$ is the areal weight of the triangular peak in the distribution. $\lambda$ is the parameter of the exponential function. $\gamma$=$\frac{1-S_{p}}{\left(1-e^{-h_{\text{m}}/\lambda}\right)}$.

Defining $\tilde{D}$=$\frac{D}{h_{\text{m}}}$, $\tilde{b}$=$\frac{b}{h_{\text{m}}}$, $\tilde{\lambda}$=$\frac{\lambda}{h_{\text{m}}}$, the evolutions of $\tilde{A}$ and $\tilde{P}$ are:
\begin{align*}
\tilde{A}=&\gamma \left[(\tilde{D}-\tilde{\lambda} -1)e^{-\frac{1}{\tilde{\lambda}}}+\tilde{\lambda} e^{-\frac{\tilde{D}}{\tilde{\lambda}}} \right]\\
&+\left\{
	\begin{array}{ll} 
		\frac{S_{p}}{3\tilde{b}^2}(\tilde{D}-1)^2(3\tilde{b}+\tilde{D}-1)& \text{if} \, \tilde{D}\in[1-\tilde{b},1] \\
		- \frac{S_{p}}{3}(\tilde{b}+3\tilde{D}-3)& \text{if} \, \tilde{D}\in[0,1-\tilde{b}],
	\end{array}
	\right.
\end{align*}
\begin{align*}
\tilde{P}=&\gamma f(\tilde{D})+\\
&\frac{4 S_p}{35 \tilde{b}^2}\left\{
	\begin{array}{lll} 
		&(1-\tilde{D})^{\frac{5}{2}} (7\tilde{b} + 2\tilde{D} - 2) & \text{if} \, \tilde{D}\in[1-\tilde{b},1] \\
        &[(1-\tilde{D})^{\frac{5}{2}} (7\tilde{b} + 2\tilde{D} - 2)- & \\
        &(1-\tilde{b}-\tilde{D})^{\frac{5}{2}} (2\tilde{b} + 2\tilde{D} - 2)] & \text{if} \, \tilde{D}\in[0,1-\tilde{b}],
	\end{array}
	\right.
\end{align*}
\noindent with:
$f(x)$=$\frac{3\sqrt{\pi}\tilde{\lambda}^{3/2}}{4e^{x/\tilde{\lambda}}} erf\left(\sqrt{\frac{1-x}{\tilde{\lambda}}}\right)+\frac{\sqrt{1-x}}{2e^{1/\tilde{\lambda}}}(2x-3\tilde{\lambda}-2)$. 
The linear asymptote of $\tilde{A}(\tilde{P})$ when $\tilde{D}$$\to$0 is $\tilde{A}$=$\left(\tilde{A}_1-\tilde{A}_2 \frac{\tilde{P}_1}{\tilde{P}_2}\right) + \frac{\tilde{A}_2}{\tilde{P}_2} \tilde{P}$, where $\tilde{A}_1$=$\gamma \left[\tilde{\lambda}-(1+\tilde{\lambda})e^{-\frac{1}{\tilde{\lambda}}}\right]-\frac{S_p}{3}(\tilde{b}-3)$, 
$\tilde{A}_2$=$\gamma \left(e^{-\frac{1}{\tilde{\lambda}}}-1\right)-S_p$, 
$\tilde{P}_1$=$\gamma \left[\frac{3\sqrt{\pi}\tilde{\lambda}^{\frac{3}{2}}}{4}erf\left(\sqrt{\frac{1}{\tilde{\lambda}}}\right)-\left(1+\frac{3\tilde{\lambda}}{2}\right)e^{-\frac{1}{\tilde{\lambda}}}\right]$+$\frac{4S_p}{35 \tilde{b}^{2}} \left(2 (1-\tilde{b})^{\frac{7}{2}}+7\tilde{b}-2\right)$ and
$\tilde{P}_2$=$\frac{3\gamma}{4} \left[(\tilde{\lambda}+2)e^{-\frac{1}{\tilde{\lambda}}}-\sqrt{\pi \tilde{\lambda}}\left(erf\left(\sqrt{\frac{1}{\tilde{\lambda}}}\right)+\sqrt{\frac{\tilde{\lambda}}{\pi}}e^{-\frac{1}{\tilde{\lambda}}}\right)\right]-\frac{2S_p}{5 \tilde{b}^{2}} \left(2 (1-\tilde{b})^{\frac{5}{2}}+5\tilde{b}-2\right)$. In Fig.~\ref{fig-linear_friction_law}B, the friction law corresponds to $\tilde{\lambda}$=0.2617, $\tilde{b}$=0.0333 and $S_p$=0.0740, obtained applying the three-parameters GA optimisation where, in $f_1$ and $f_2$, $a$ and $b$ are the slope and intercept of the above-defined asymptote of $\tilde{A}(\tilde{P})$. The list of heights is determined from the continuous height distribution by dividing the interval $[0,h_{\text{m}}]$ into 64 bins of equal 1/64 probability in $\phi(h)$, and defining $h_i$ as the center of the $i^{\text{th}}$ interval.


\section*{Acknowledgements}
The authors are indebted to the Carnot institute Ingénierie@Lyon, labelled by the French National Research Agency (ANR), for its support and funding. We thank S. Zerrouk (mould manufacturing), A. C. Pontes Rodrigues (logistics), M. Guibert (instrumentation) and N. Morgado and M. Catherin (sample metallisation) for their help.

\section*{Author contribution}
L.F., D.D. and J.S. designed research. L.F. performed optimisations and analyzed the numerical data. D.G.K. performed experiments and analyzed the experimental data. J.S. performed analytical modelling. L.F. and J.S. wrote the paper. All authors reviewed the manuscript.
To whom correspondence should be addressed. E-mail: julien.scheibert@cnrs.fr \& li.fu@ec-lyon.fr



\clearpage
\section*{Supplementary Information}
\renewcommand{\thefigure}{S\arabic{figure}}
\renewcommand{\thetable}{S\arabic{table}}
\setcounter{figure}{0}
\setcounter{table}{0}


\section{Systematic benchmark of optimisation strategies}

To guide our choice of a suitable optimisation-based design method, we perform a controlled comparison between four representative search algorithms, three model-free metaheuristics (Simulated Annealing, SA; Genetic Algorithm, GA; Particle Swarm optimisation, PSO) and one probabilistic model-based (Bayesian optimisation, BO). We use them to solve the same inverse design problem (a power friction law, denoted as Case 2 in the main text) under matched conditions. To compare optimisation strategies and not implementation artifacts, all methods share a common objective function, common parameter bounds, and a common target set for the power-law exponent.

Methodologically, the four optimisers are selected because they span complementary search behaviors. SA emphasizes thermodynamic hill-climbing with controlled uphill moves; GA provides population-based recombination and mutation; PSO captures collective swarm dynamics balancing personal and social learning; BO offers sample-efficient surrogate-guided exploration/exploitation. 

The benchmark is built around three principles. First, budget parity: each algorithm is run under a comparable computational budget so that differences in performance reflect search efficiency rather than brute-force sampling. In practice, we allocate a fixed number of fitness evaluations (60000 times) and map it to each optimiser's native iteration scheme (e.g., generations for GA, swarm iterations for PSO, calls for BO). Second, statistical robustness: each method is repeated over 30 independent runs per target, allowing us to quantify stochastic variability from random initialisation and trajectory noise, and to report mean/standard deviation rather than single-run outcomes. Third, multi-criteria assessment: beyond the final objective value, we report exponent error, $R^2$, runtime, and global fitness for all targets, so that solution quality and computational efficiency are jointly assessed.

To ensure methodological fairness across algorithms (including methods that do not natively support multi-objective optimisation), we convert the original two-objective problem of Case 2 into a single scalar fitness. For a candidate height vector, we compute the fitted exponent ($k$) and goodness of fit ($R^2$), then define

\[
f_k=\frac{1}{1+\left|k-k_{\mathrm{target}}\right|/k_{\mathrm{target}}},\qquad
f_{R^2}=\frac{1}{1+(1-R^2)},
\]
and optimise

\[
f_\text{scalar}=0.5 f_k+0.5 f_{R^2}.
\]

Key hyperparameters are selected to reflect standard strong baselines. SA uses $T_0=100$, cooling rate 0.9995, $T_{\min}=10^{-3}$, and adaptive cooling/neighbour perturbation. GA uses population size 500, crossover probability 0.4, polynomial bounded mutation (mutation probability 0.9; mutation distribution index 0.5) and tournament selection (size 3). PSO uses 500 particles with inertia $w=0.7$, cognitive/social coefficients $c_1=c_2=1.5$, and velocity clipping for stability. BO uses Gaussian Process-based optimisation with Expected Improvement, $n_{\text{calls}}= 200$, $n_{\text{initial}}= 20$, $\xi=0.01$, and a Matérn-kernel surrogate.

\begin{figure}[t!]
\centering
\includegraphics[width=0.77\textwidth]{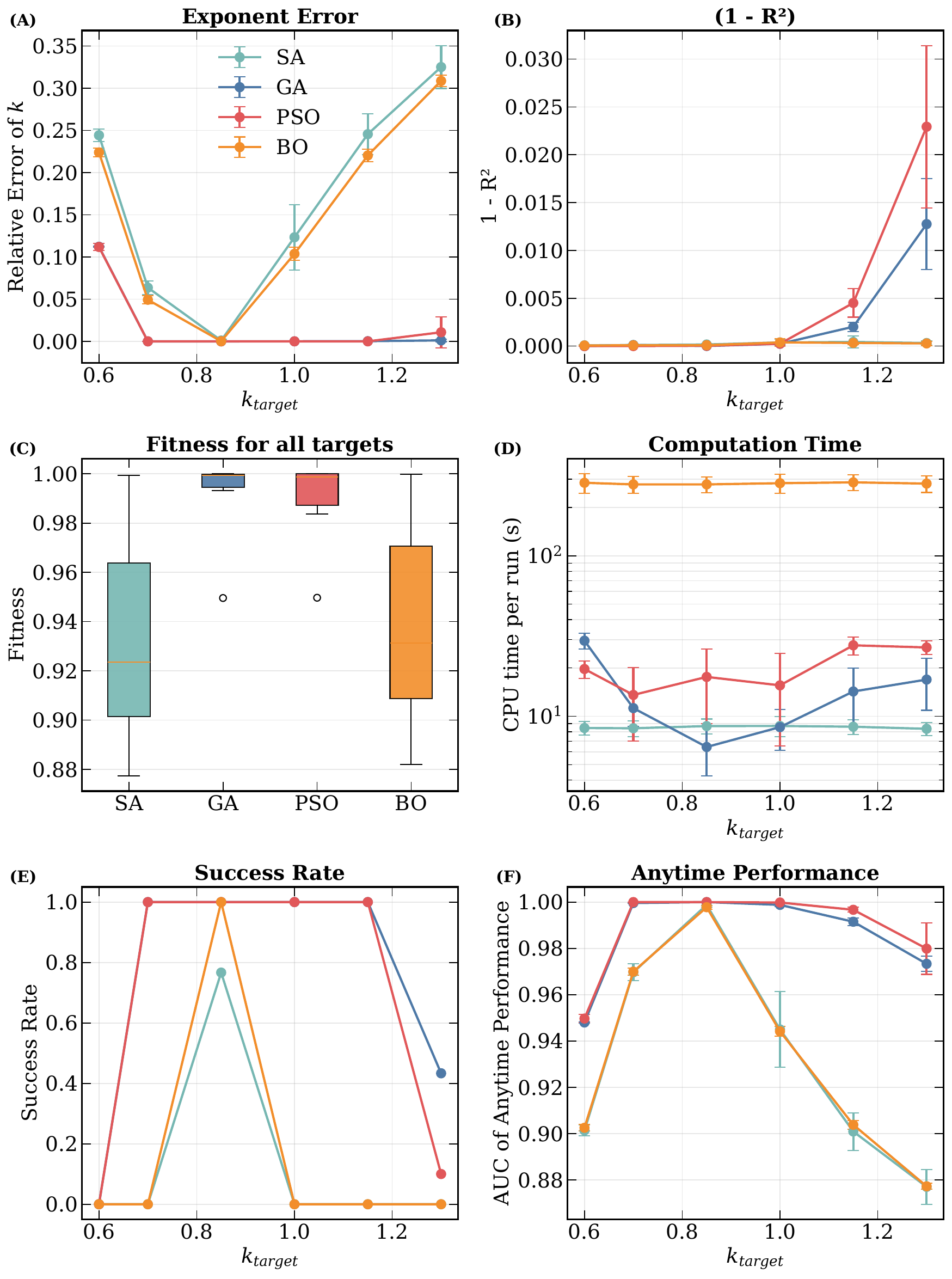}
\caption{Cross-target comparison of the four optimisation strategies (SA, GA, PSO, BO, see legend in panel A for colors) for Case 2 (power friction law conception). Relative error on exponent $k$ (A), 1 minus the goodness of fit $R^2$ (B), computation time (D), success rate (E) and area under the curve (AUC) of anytime performance (F) as functions of the target exponent, $k_\text{target}$. (C): distributions of the fitness across all $k_\text{target}$, in a boxplot representation (the orange central line is the median, the box spans the interquartile range (IQR), the whiskers span all points that are out of the box but closer to it than 1.5$\times$IQR, and open disks are outliers).}
\label{fig-cross_target_comparison}
\end{figure}

The benchmark results, summarized in Fig.~\ref{fig-cross_target_comparison}, evaluate the algorithms based on final solution quality, computational cost, success rate, and anytime performance. 

We can first disqualify SA and BO. The fitness distribution (Fig.~\ref{fig-cross_target_comparison}C) reveals that both SA and BO suffer from severe variance and frequently converge to sub-optimal solutions, with fitness scores dropping as low as 0.88 (see panels A and B for errors on each individual objective). Apart from the particular case of $k_{\text{target}}$=0.85, their success rate is extremely low, and their anytime performance is not competitive (Fig.~\ref{fig-cross_target_comparison}F). Furthermore, BO exhibits a significant computational overhead (Fig.~\ref{fig-cross_target_comparison}D) due to the use of a Gaussian Process surrogate model. Since the evaluation of our contact mechanics model is itself computationally inexpensive, the surrogate model provides no meaningful benefit in reducing the number of objective function evaluations; instead, it introduces a substantial bookkeeping cost, requiring CPU times two orders of magnitude higher than the other methods. Consequently, neither SA nor BO are viable candidates for our metainterface conception. 

The comparison then narrows to GA and PSO. While both algorithms successfully handle the optimisation tasks with high success rates (Fig.~\ref{fig-cross_target_comparison}E), GA consistently achieves the highest fitness with an exceptionally tight distribution effectively at 1.00 (Fig.~\ref{fig-cross_target_comparison}C). By providing the tightest convergence stability, maintaining near-perfect success rates (Fig.~\ref{fig-cross_target_comparison}E), and operating with higher computational efficiency than its closest competitor (Fig.~\ref{fig-cross_target_comparison}D), GA emerges as the most robust and practical candidate algorithm for this study.

\newpage

\section{Optimisation metrics of the GA-based inverse design}

To characterize statistically the properties of the solutions provided by the Genetic Algorithm, all optimisations have been performed for at least 30 independent runs per target law. Below, we discuss typical results that illustrate convergence, reproducibility, scalability of the inverse design process, as well as the quality and diversity of the obtained solutions.

\subsection{Convergence stability and reproducibility}

\begin{figure}[h!]
     \centering
     \includegraphics[width=0.75\linewidth]{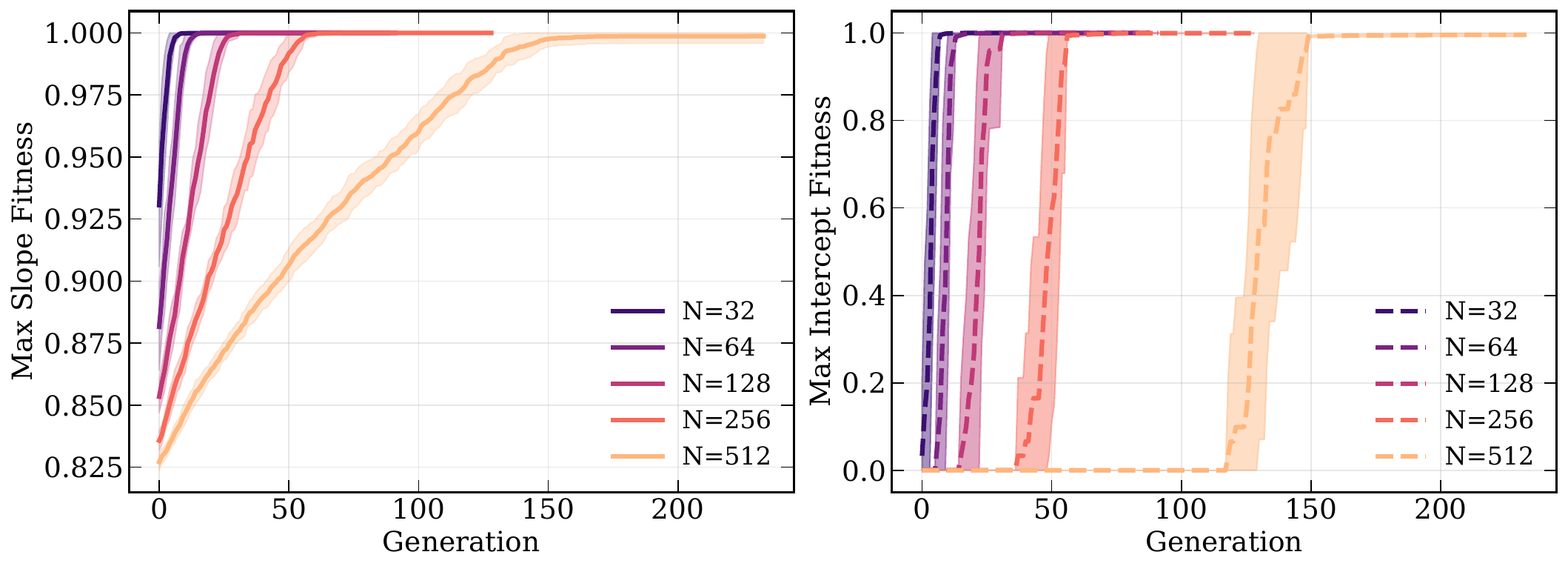}
     \includegraphics[width=0.45\linewidth]{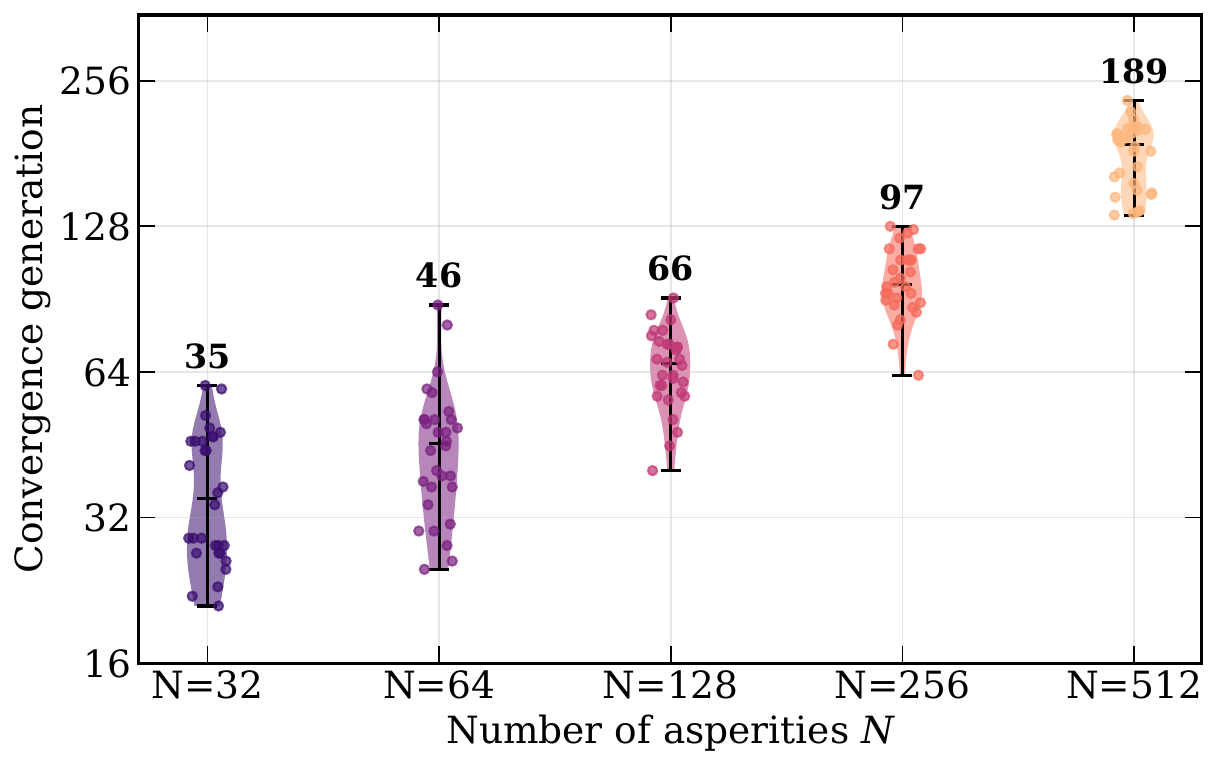}
        \caption{Convergence curves for the optimisation of a proportional law with target slope $a_{\text{target}}$=1.39, for various numbers of asperities, $N$ (see legend). Upper left (resp. upper right): evolution of the maximum slope fitness (resp. maximum intercept fitness) within a generation as a function of the index of the generation. Each curve is the average of 30 optimisation runs, while the shaded area contains all 30 curves. Lower panel: convergence generation, defined as the first generation for which both fitness values vary by less than $10^{-5}$ over the next 50 generations, as a function of $N$. Median values are indicated.}
\label{fig:linear_Effect_N}
\end{figure}

\begin{figure}[h!]
     \centering
     \includegraphics[width=0.75\linewidth]{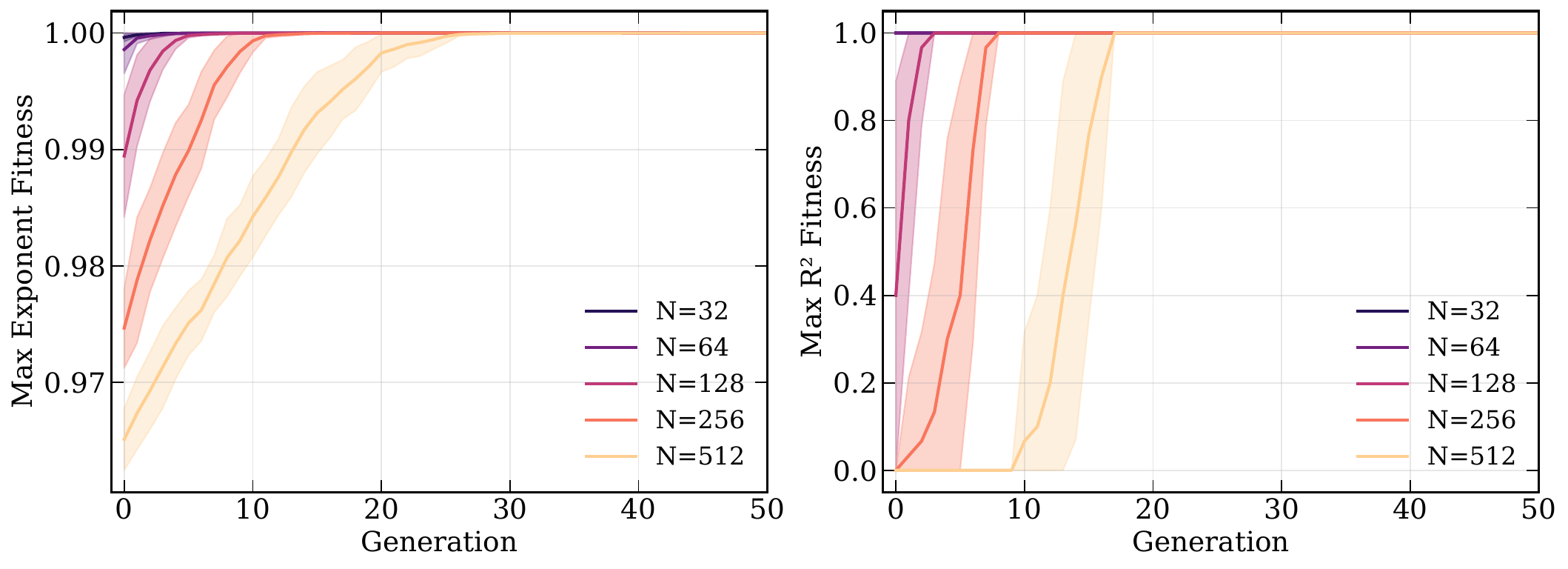}
     \includegraphics[width=0.45\linewidth]{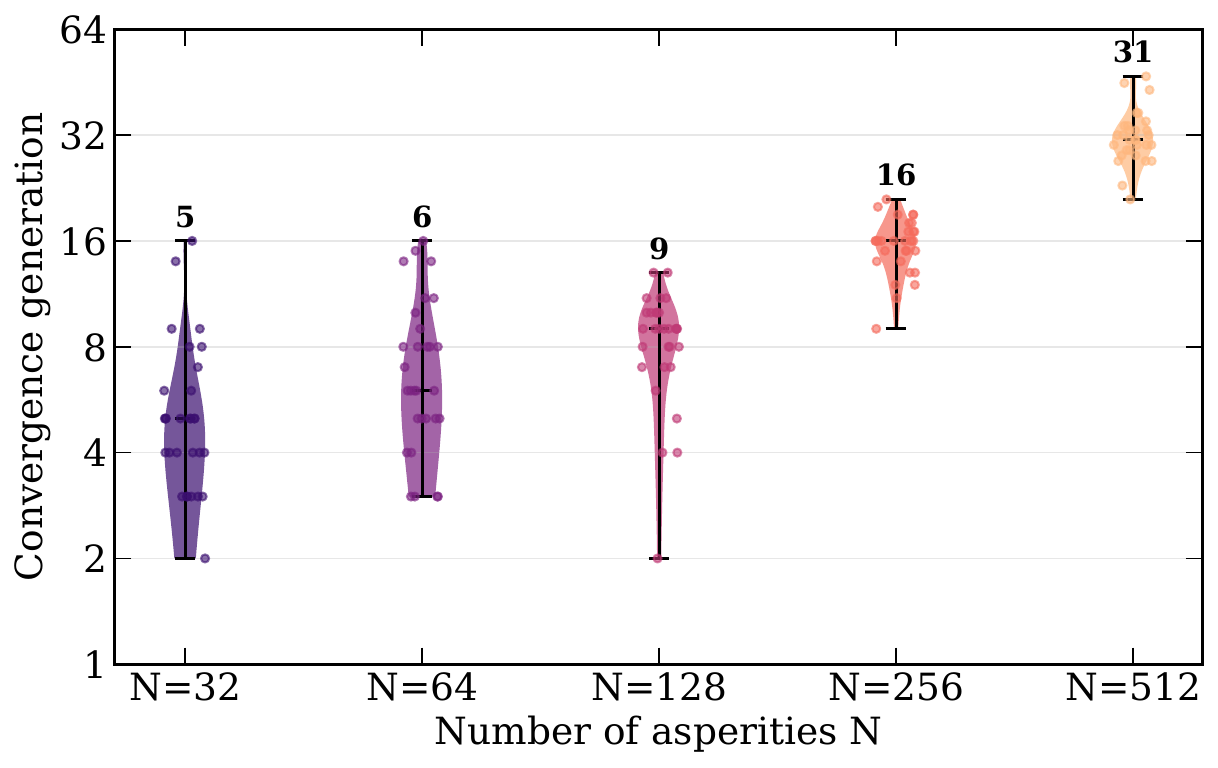}
        \caption{Convergence curves for the optimisation of a power law with target exponent $k_{\text{target}}$=0.85, for various numbers of asperities, $N$ (see legend). Upper left (resp. upper right): evolution of the maximum exponent fitness (resp. maximum goodness-of-fit fitness) within a generation as a function of the index of the generation. Each curve is the average of 30 optimisation runs, while the shaded area contains all 30 curves. Lower panel: convergence generation, defined as the first generation for which both fitness values vary by less than $10^{-5}$ over the next 50 generations, as a function of $N$. Median values are indicated.}
\label{fig:power_Effect_N}
\end{figure}

\begin{figure}[h!]
     \centering
     \includegraphics[width=0.86\linewidth]{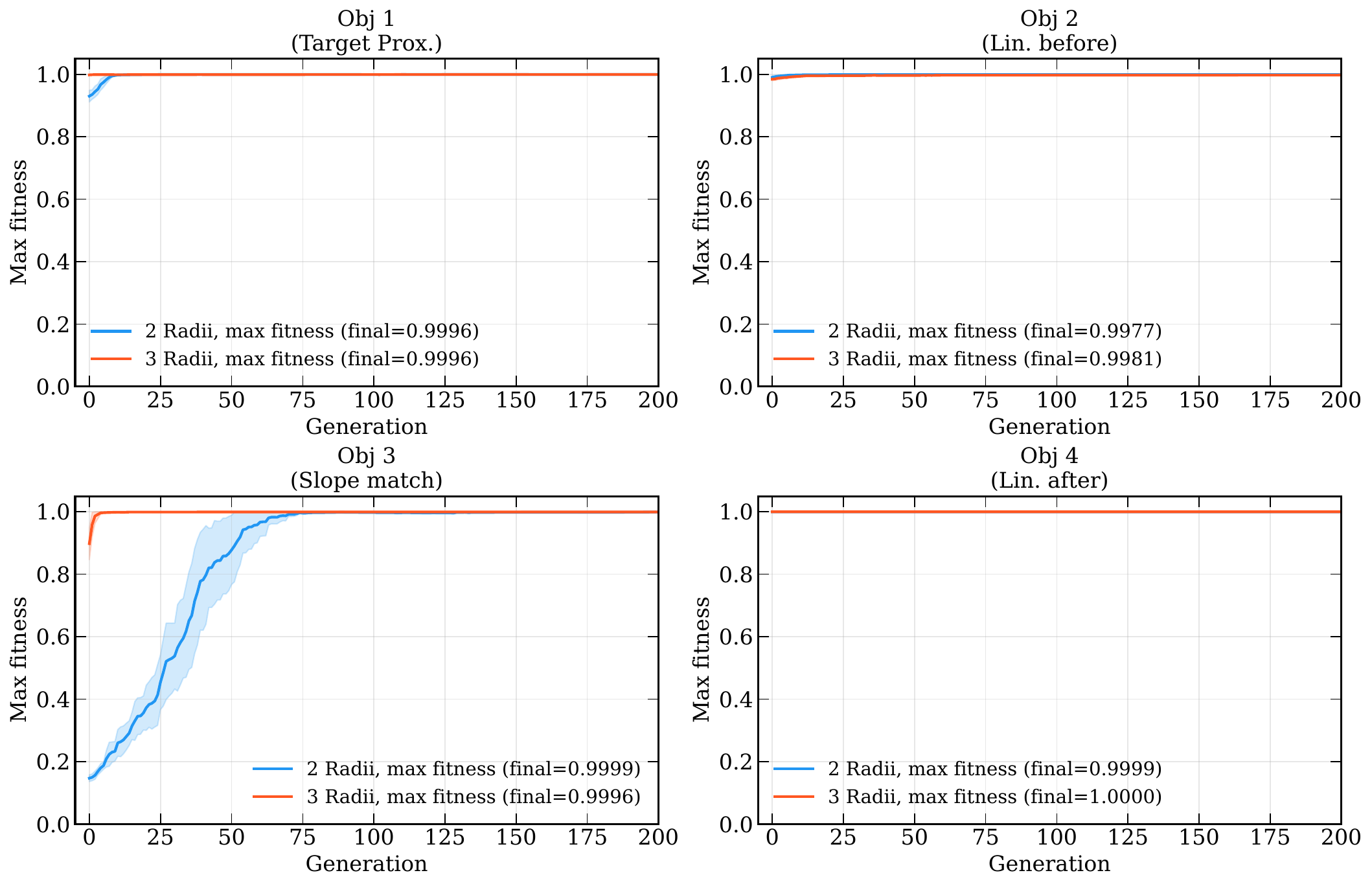}
        \caption{Convergence curves for the optimisation of a bilinear friction law with target point (0.06, 0.12), target second segment slope $m_\mathrm{2, target}=1.2$ and $N=64$, using 2 ($R$ and $3R$, blue curves) or 3 ($0.5R$, $R$ and $1.5R$, red curves) possible values of the radius. Each panel represents one maximum fitness value of the four objectives. From top left to bottom right: proximity to the target point; goodness-of-linear-fit in the first branch; accuracy of the slope of the second branch; goodness-of-linear-fit in the second branch. Each curve is the average of 30 optimisation runs, while the shaded area contains all 30 curves.}
\label{fig:2branches_convergence_comparaison_nb_R}
\end{figure}

Typical average convergence curves are shown as lines in Figs.~\ref{fig:linear_Effect_N}, \ref{fig:power_Effect_N} and \ref{fig:2branches_convergence_comparaison_nb_R} for a proportional law (see Case 1 in the main text), a power law (Case 2) and a bilinear law (Case 3), respectively. In all cases, the optimisation converges well within 1000 generations.

Reproducibility of the search process is demonstrated by the narrowness of the colored bands that surround the average curves in Figs.~\ref{fig:linear_Effect_N}, \ref{fig:power_Effect_N} and \ref{fig:2branches_convergence_comparaison_nb_R}, and that contain all 30 individual convergence curves.

\begin{figure}[h!]
     \centering
     \includegraphics[width=0.49\linewidth]{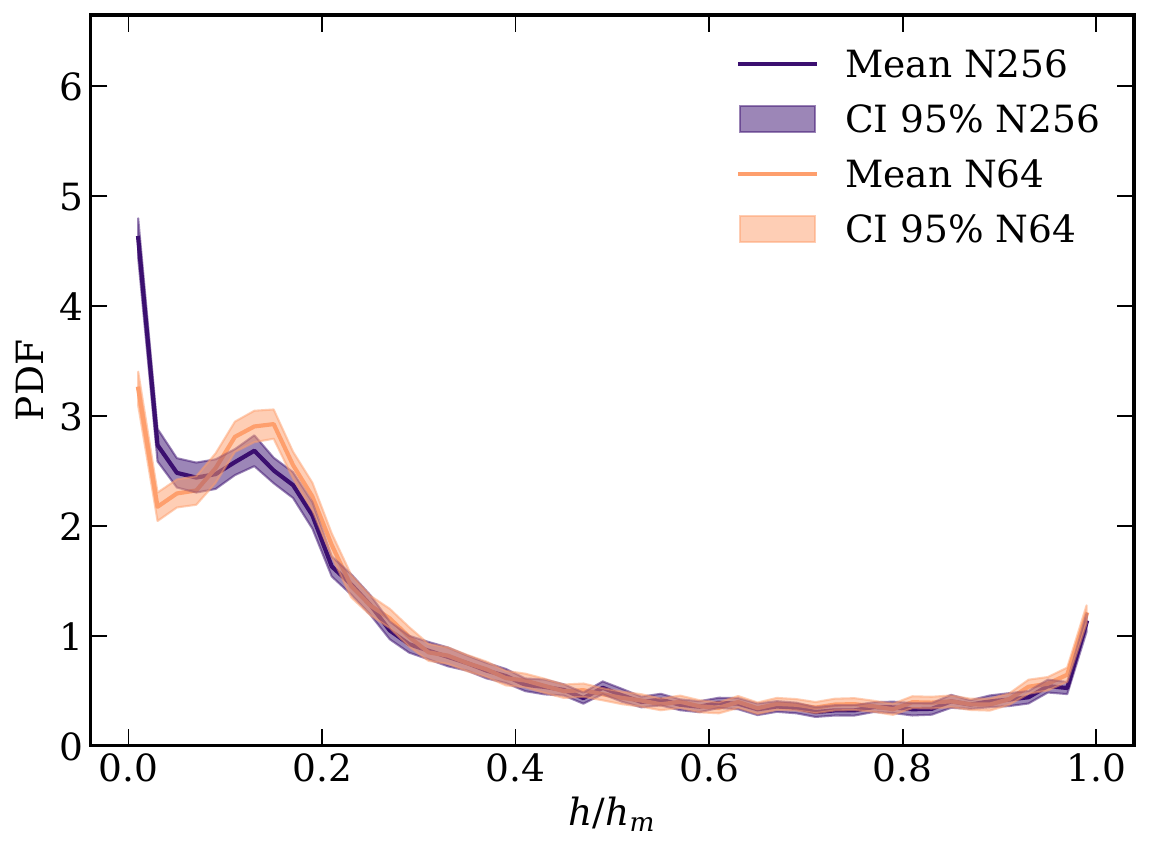}
     \includegraphics[width=0.49\linewidth]{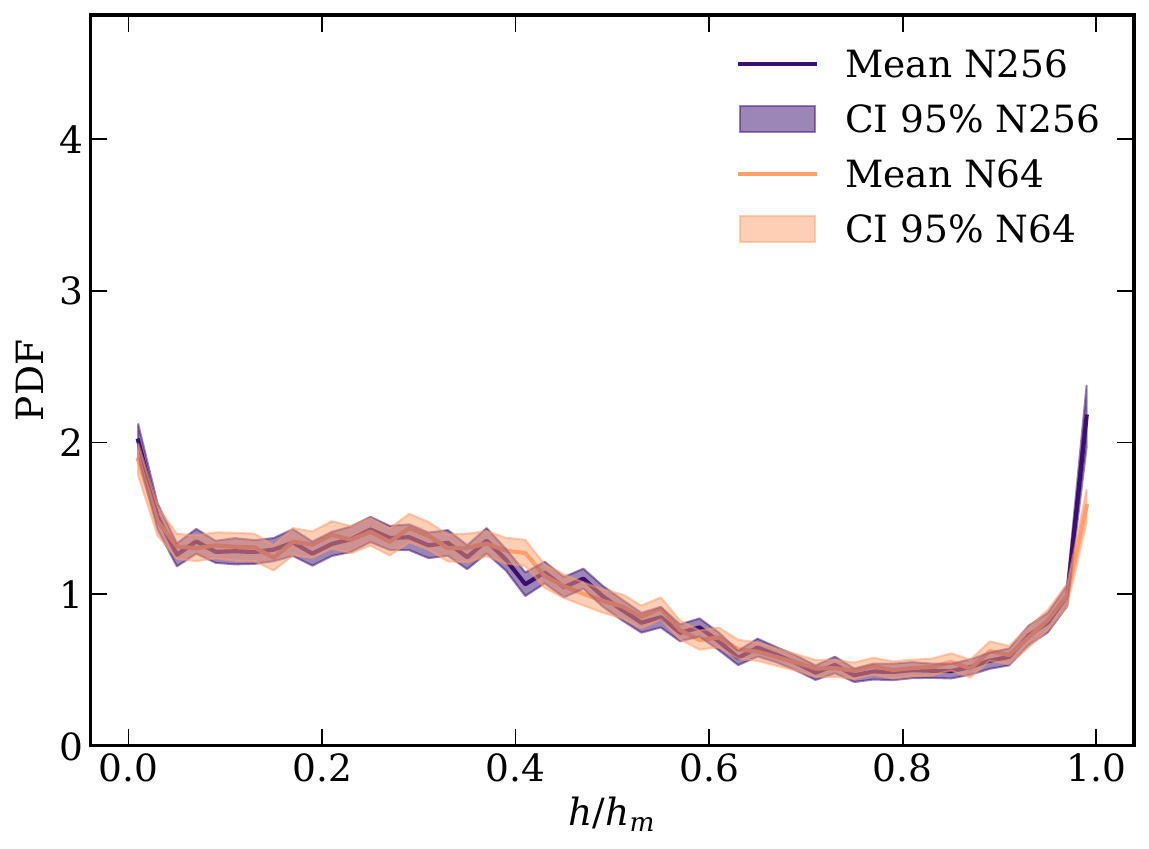}
        \caption{Optimised probability density functions (PDFs) of the asperity heights for Case 1 ($a_\text{target}=1.39$, left) and Case 2 ($k_\text{target}=0.85$, right). Results from 100 and 400 optimisation runs are overlaid for $N=256$ (purple) and $N=64$ (red), respectively. Mean values (solid lines) and 95\% confidence intervals (shaded bands) are shown.}
\label{fig:histogram_Effect_N}
\end{figure}

Reproducibility of the solutions is then illustrated, for Cases 1 and 2, by the narrowness of the colored bands in Fig.~\ref{fig:histogram_Effect_N}, which illustrate the 95\% confidence interval of the probability distribution function (PDF) of the asperity heights.

\subsection{Scalability}

The evolution of the convergence rate as the number of design variables, $N$, increases is illustrated in Figs.~\ref{fig:linear_Effect_N} and \ref{fig:power_Effect_N}, where the convergence curves are compared for $N$ varying from 32 to 512. As expected, a larger number of design variables slows convergence. As shown in the lower panels, in the explored range where $N$ varies by a factor of 8, the median generation index at which convergence is observed increases by less than a factor of 3.7 (resp. 6.2) for the proportional law of Fig.~\ref{fig:linear_Effect_N} (resp. the power law of Fig.~\ref{fig:power_Effect_N}).

\subsection{Optimisation trade-offs}

\begin{figure}[b!]
\centering
\includegraphics[width=0.49\textwidth]{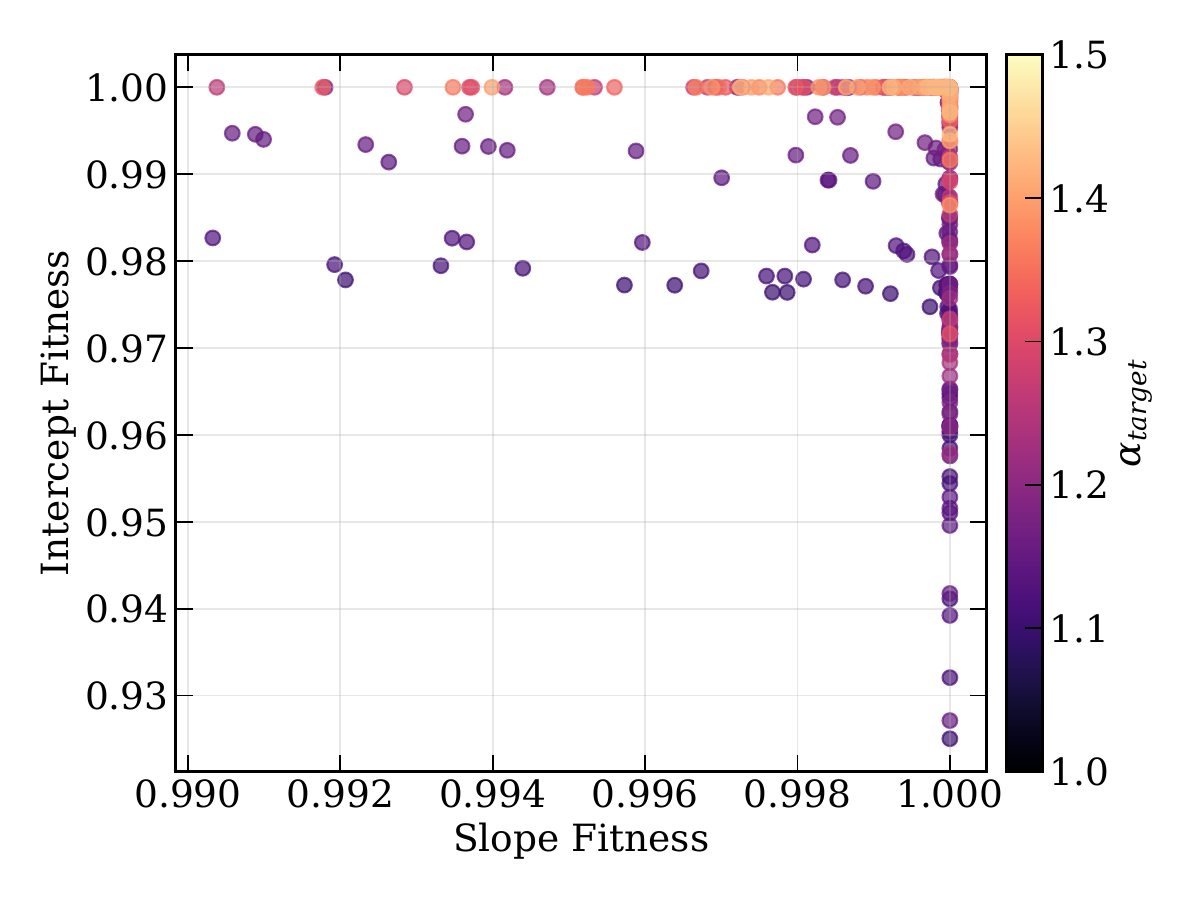}
\includegraphics[width=0.49\textwidth]{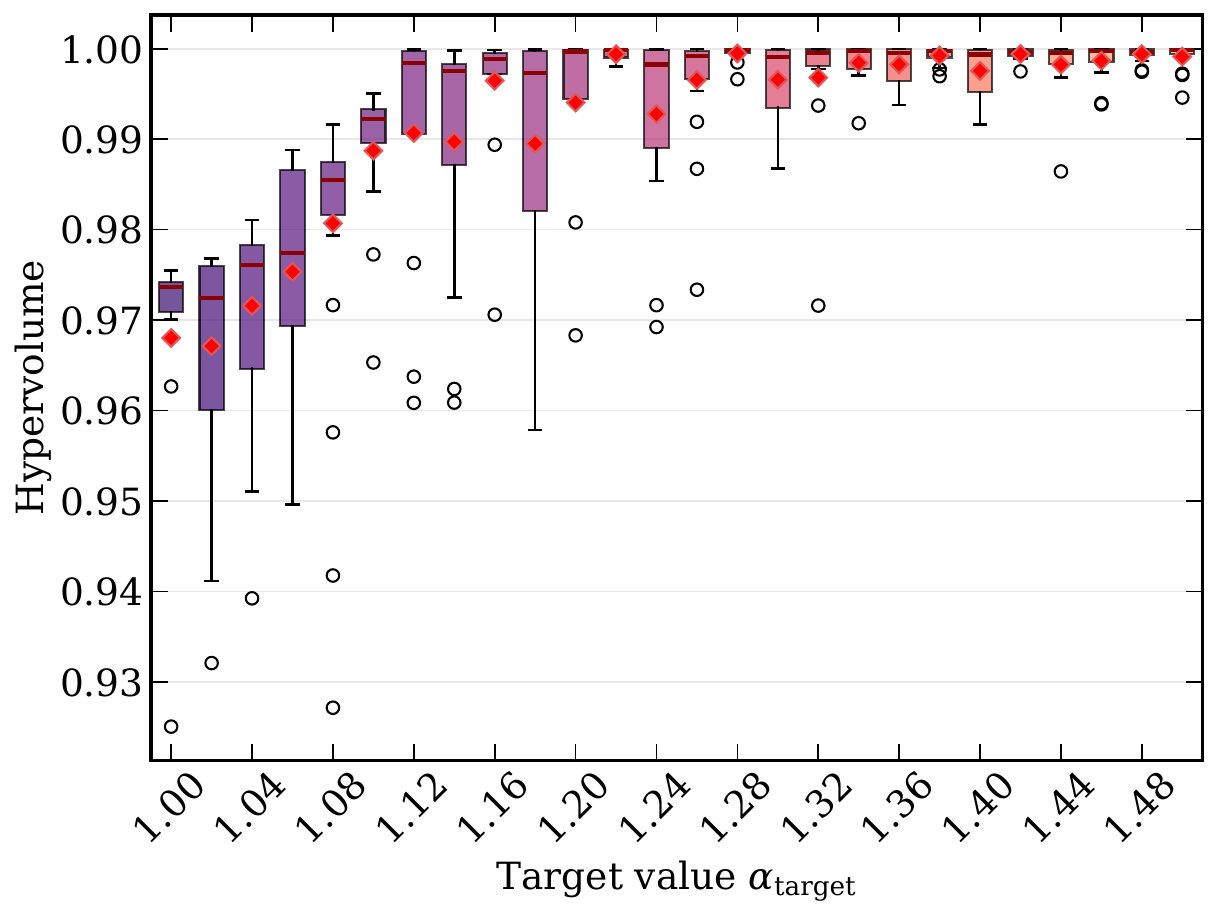}
\caption{Left: Global Pareto fronts obtained for different target exponents $a_{target}$ in the two-objectives optimisation of linear laws. Each Pareto front is aggregated from 30 optimisation runs. Right: distribution of hypervolumes as a function of $a_{target}$ in a boxplot representation: the central line is the median, the red point is the mean, the box spans the interquartile range (IQR), the whiskers span all points that are out of the box but closer to it than 1.5$\times$IQR, and open disks are outliers.} 
\label{fig:linear_law_pf}
\end{figure}

\begin{figure}[t!]
    \centering
    \includegraphics[width=0.49\linewidth]{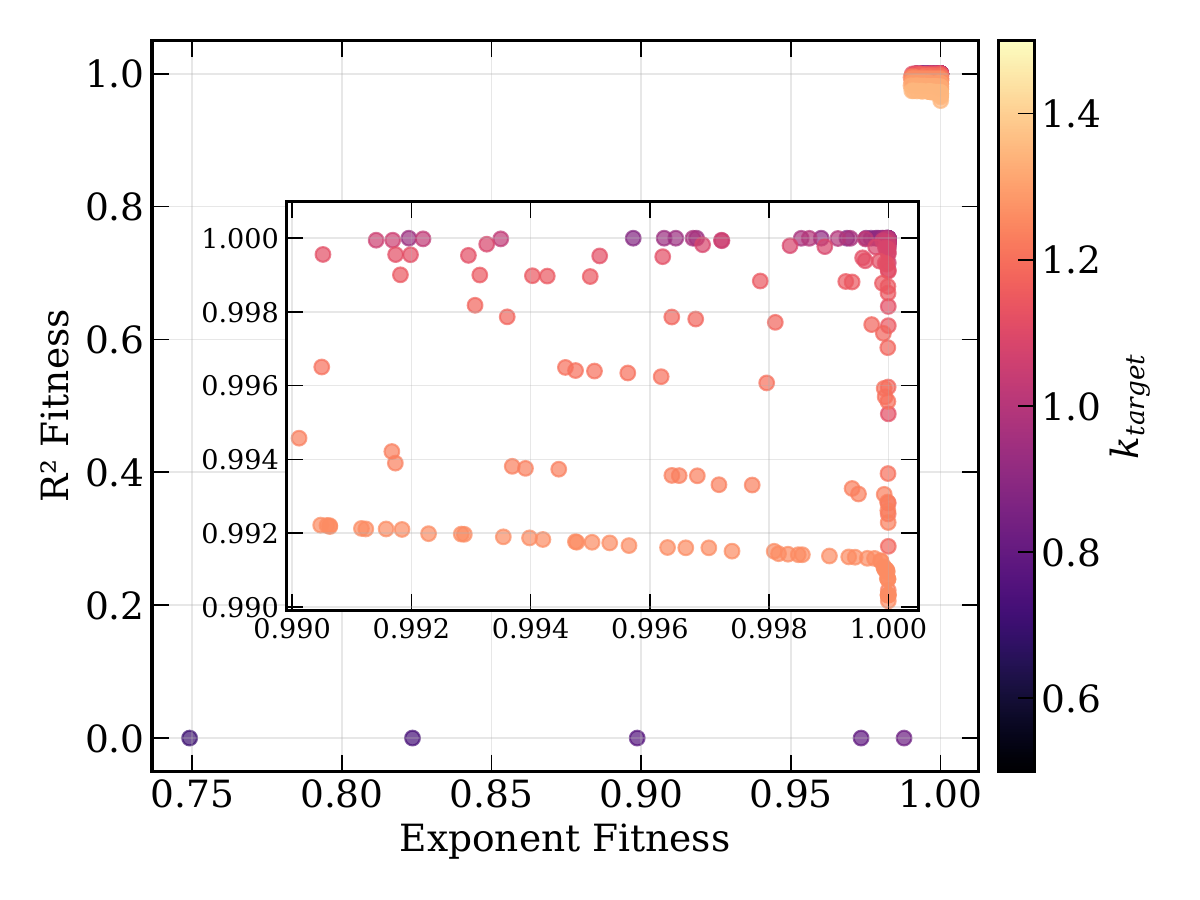}
    \includegraphics[width=0.48\linewidth]{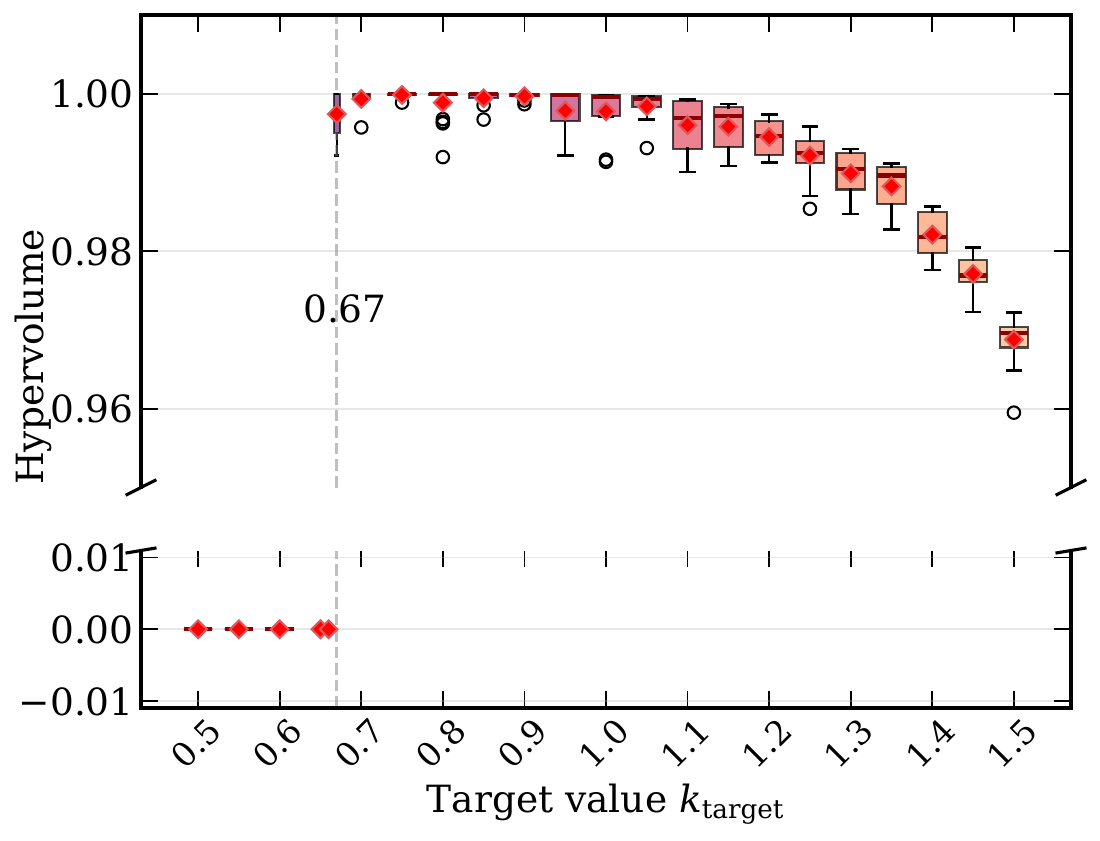}
    \caption{Left: Global Pareto fronts obtained for different target exponents $k_{\text{target}}$ in the two-objectives optimisation of power laws. Each Pareto front is aggregated from 30 optimisation runs. Inset: zoom in the region close to the ideal solution $(1,1)$.  Right: distribution of hypervolumes as a function of $k_{\text{target}}$ in a boxplot representation similar to that in Fig.~\ref{fig:linear_law_pf}.}
    \label{fig:power_law_pf}
\end{figure}

The optimisations performed in this work are multi-objective. For instance, for Cases~1 and~2, NSGA-II minimises two objectives simultaneously (for Case~1, the slope fitness and the intercept fitness; for Case~2, the exponent fitness and the goodness-of-fit fitness). Both objectives are normalised so that a perfect solution would correspond to a value of~1 for both fitnesses. The natural representation of the trade-off between objectives is the Pareto front in the two-dimensional objective space. Figures~\ref{fig:linear_law_pf} and~\ref{fig:power_law_pf} display the global Pareto fronts aggregated from the 30 independent optimisation runs for each target value. A front concentrated near the ideal point $(1,1)$ indicates that no meaningful trade-off exists: both objectives can be simultaneously optimised to near-perfect values. Conversely, a front that extends away from $(1,1)$ reveals a genuine tension between the two objectives — improving one necessarily degrades the other. 
For Case~1, the Pareto fronts are concentrated near $(1,1)$ for all accessible target slopes ($a_{\text{target}} \ge 1.12$), indicating that slope accuracy and intercept minimisation can be simultaneously achieved with no meaningful compromise. The same holds for Case~2 within the accessible exponent range $k_{\text{target}} \in [2/3, 1]$: the fronts remain close to $(1,1)$, showing that the exponent accuracy and the goodness-of-fit are not in genuine conflict. Only as $a_{\text{target}} < 1.12$ or $k_{\text{target}} >1 $ do the fronts depart progressively from the ideal point, signalling that a true trade-off between the two objectives emerges and that the target becomes intrinsically harder to achieve.

\begin{figure}[ht!]
     \centering
     \includegraphics[width=0.49\linewidth]{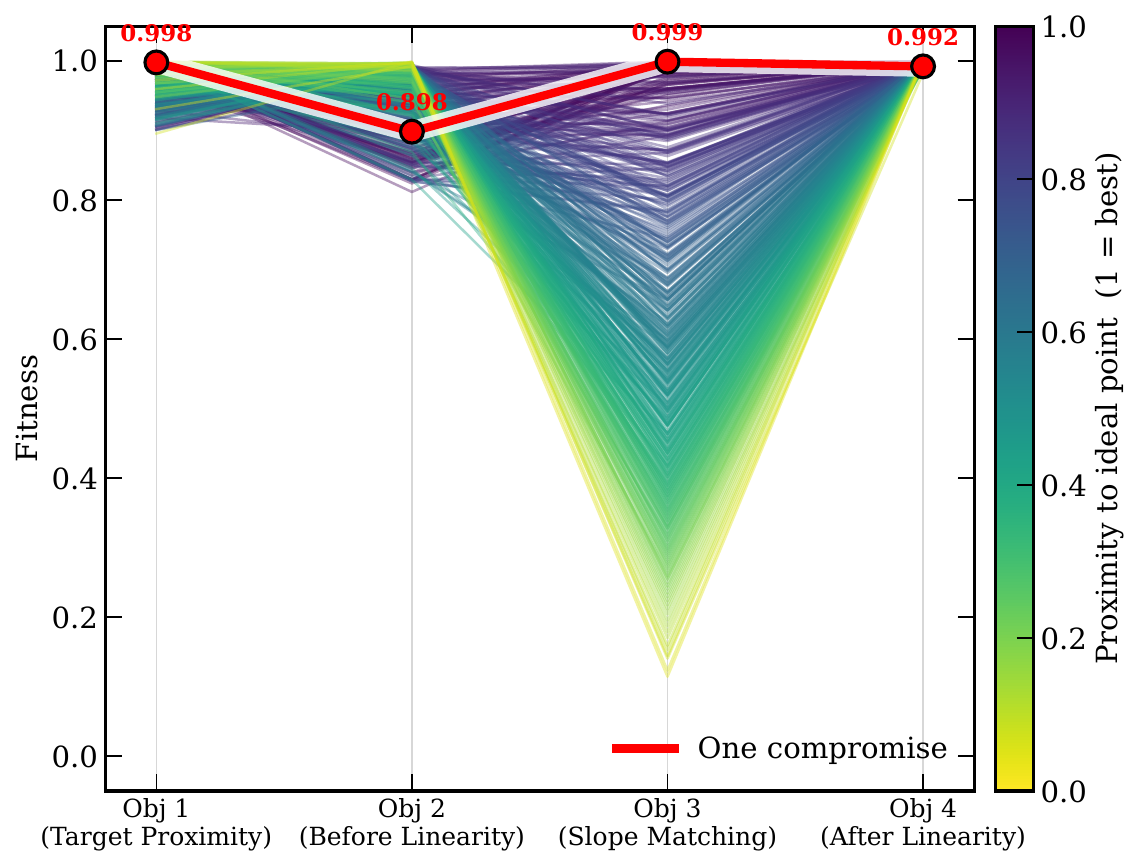}
     \includegraphics[width=0.49\linewidth]{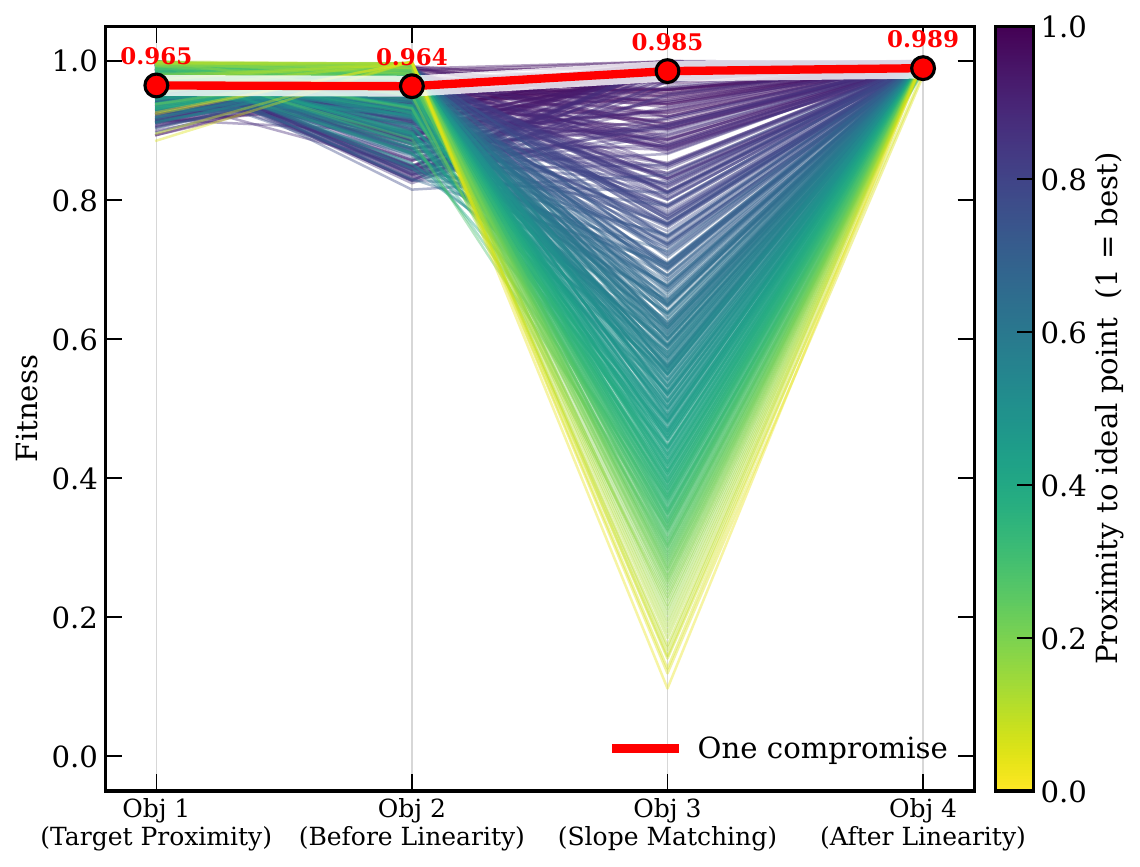}
     \includegraphics[width=0.5\linewidth]{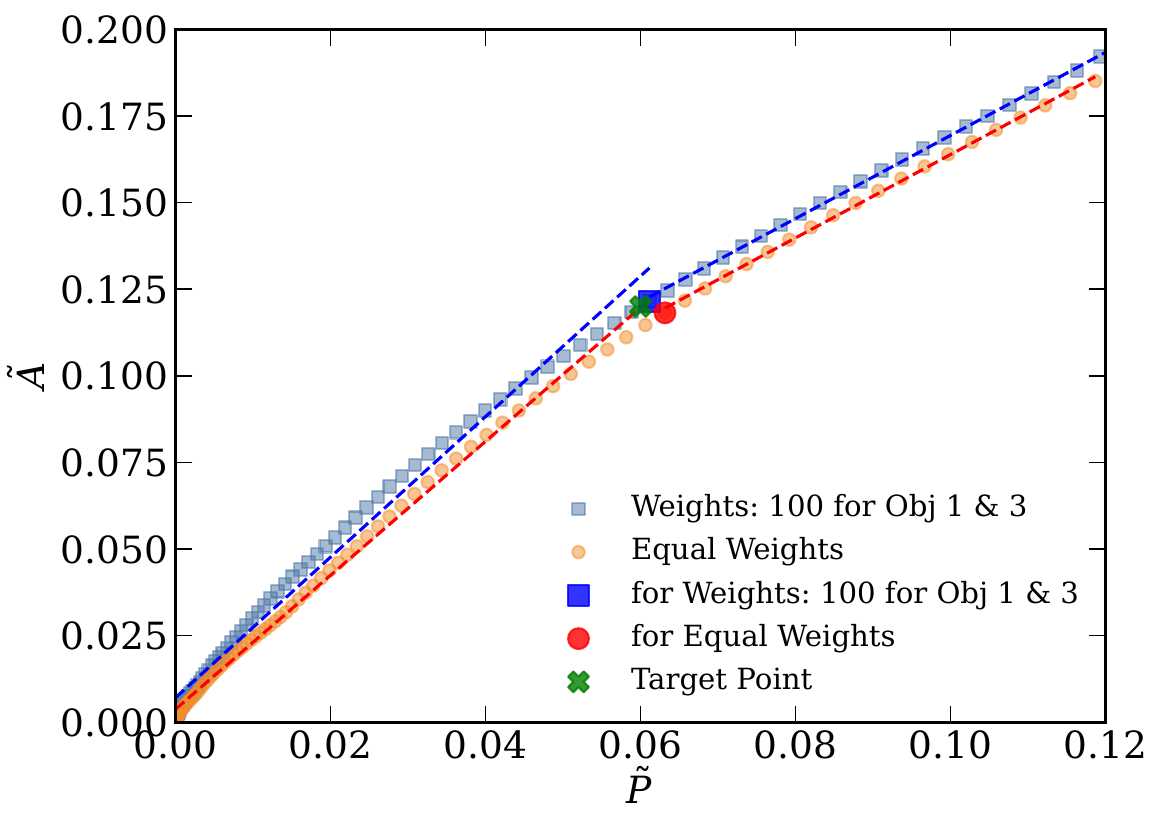}
        \caption{Parallel coordinates plot of the Pareto front for Case 3 (bilinear friction law, target point (0.06, 0.12), target second segment slope $m_\mathrm{2, target}=1.2$). Each line represents a solution on the aggregated Pareto front, and the color indicates its weighted distance to the ideal point (1,1,1,1). With weight ratios of (100:1:100:1) (upper left) or (1:1:1:1) (upper right) for each objective, the best solution is highlighted in red. The lower panel shows the comparison of the two optimisation results with different weight ratios. Bigger symbols represent the closest points to the target point. Dashed lines represent the linear fits of the first and second segments of each solution.} 
 \label{fig:parallel_coordinates_case3}
\end{figure}

For Case~3 (bilinear friction law), NSGA-II simultaneously optimises four objectives: the proximity to the target crossover point, the slope accuracy of the second segment, and the goodness-of-fit of each segment. With four objectives, a Pareto front cannot be visualised in a plane, and we instead use a parallel coordinates plot (Fig.~\ref{fig:parallel_coordinates_case3}), in which each solution on the aggregated Pareto front is represented by a polyline and the colour encodes the weighted distance to the ideal point $(1,1,1,1)$. This representation makes the trade-offs explicit: solutions with high target proximity tend to have slightly lower goodness-of-fit in the first segment, and vice versa. 
To select a unique best solution from this front, different weighting strategies are possible. Here we illustrate two strategies. In the first, the two accuracy objectives (target proximity and slope matching) are given priority over the goodness-of-fit objectives, with a weight ratio of $(100:1:100:1)$ (Fig.~\ref{fig:parallel_coordinates_case3}, upper left). In the second, all four objectives are weighted equally at $(1:1:1:1)$ (Fig.~\ref{fig:parallel_coordinates_case3}, upper right). The corresponding optimised friction laws for both selected solutions are compared in the lower panel of Fig.~\ref{fig:parallel_coordinates_case3}, showing that the practical difference between the two selection strategies is small. In our experimental validation of Fig.~4B, we have given priority to an accurate reaching of the prescribed target point and therefore have adopted the weighting strategy $(100:1:100:1)$. Doing so, we accept that the initial segment has a slightly degraded goodness of linear fit.

\subsection{Solution quality}

The Pareto fronts in the left panels of Figs.~\ref{fig:linear_law_pf} and \ref{fig:power_law_pf} can be further used to build the hypervolume (HV) metric~\cite{deb2002}. HV is useful to identify the set of accessible target values (those for which the optimisation reliably converges to a near-optimal solution) and assess solution quality. The HV of a Pareto front is the area in objective space that is dominated by the front and bounded by a chosen reference point (here $(0, 0)$); with both objectives normalised to $[0;1]$, the maximum possible HV is 1, achieved  when the front reaches the ideal  point $(1, 1)$. A value close to 0 therefore indicates that at least one objective has not been optimised at all, i.e. the optimisation has failed for this target. Because each of the 30 independent runs yields its own Pareto front and hence its own HV, the box plot shows the distribution of those 30 values (see right panels  of Figs.~\ref{fig:linear_law_pf} and \ref{fig:power_law_pf}). A narrow box signals that all runs converge consistently to the same quality of solution, regardless of the random initialisation (a hallmark of a robust, well-defined fitness landscape).  Conversely, a wide boxplot means that some runs reach near-optimal solutions while others stagnate, reflecting a rugged or degenerate fitness landscape where the GA outcome is sensitive to the starting population.

For the proportional law (Fig.~\ref{fig:linear_law_pf}, right), the maximum HV is close to 1 only for $a_{\text{target}}>1.12$. Below 1.12, no relevant solution can be found. The boxplot of HV values becomes narrower when increasing $a_{\text{target}}$, indicating that designing proportional laws with higher slopes is easier and more reproducible.

For the power law (Fig.~\ref{fig:power_law_pf}, right), the maximum HV value vanishes for $k_{\text{target}} < 2/3$, suggesting that no solution exists in this range. Instead, HV close to 1 with a narrow boxplot is found for $k_{\text{target}} \in [2/3, 1]$, showing that near-perfect solutions are robustly identified. For $k_{\text{target}} > 1$, the HV value decreases and the boxplots broaden progressively as $k_{\text{target}}$ increases, indicating that designing power laws with exponents larger than 1 is both harder and less reproducible.

In practice, within the range of parameters where HV does not vanish, one still further needs to choose arbitrary criteria on the fitness values to decide whether a solution is deemed sufficiently good. This is what we have done to define the ranges of $a_{\text{target}}$ (Case 1) and $k_{\text{target}}$ (Case 2) for which the optimisation is considered successful.

\begin{figure}[h!]
\centering
\includegraphics[width=0.7\textwidth]{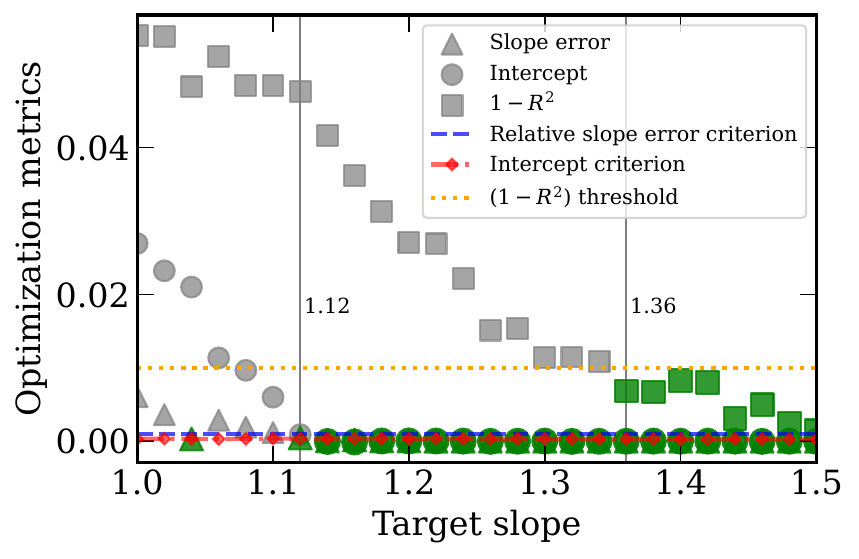}
\caption{Optimisation metrics related to the slope $a$ and intercept $b$ vs the target slope $a_\text{target}$ for Case 1 (proportional friction laws). A relative slope error ($|a-a_\text{target}|/a_\text{target}$) less than 0.1\% (dashed blue line) is achieved for all $a_{\text{target}}>1.12$. For the same range of slopes, the intercept is also minimal (less than 0.1\% of $a_{\text{target}}\tilde{P}_{\text{max}}$, dashed red line).
A good proportionality over the whole range $[0,P_{\text{max}}]$, not used as an optimisation criterion but quantified here by a goodness of proportional fit $R^2>0.99$ (dashed orange line), is obtained for all target slopes above $a_{\text{min}}=1.36$. Green symbols represent admissible target slope values according to each criterion.} 
\label{fig-case_1_metrics}
\end{figure}

For linear laws, Fig.~\ref{fig-case_1_metrics} quantifies the quality of the best solution for each target using three independent criteria: (i) relative slope error $|a - a_\text{target}|/a_\text{target} < 0.1\%$ (slope accuracy, directly optimised), (ii) intercept $|b| < 0.1\%\,a_\text{target}\tilde{P}_\text{max}$ (proportionality quality, also directly optimised), and (iii) global goodness of proportional fit $R^2 > 0.99$ (used here as an independent check of proportionality quality over the entire load range $[0, P_\text{max}]$, not enforced during optimisation). Criteria (i) and (ii) are satisfied for all $a_\text{target} > 1.12$. Criterion (iii) is more demanding: it is satisfied only for $a_\text{target} \geq a_\text{min} = 1.36$. This lower bound reflects a physical constraint arising from contact mechanics: below this value, the non-linearity inherent to Hertzian asperity contacts at low $P$ cannot be compensated by any choice of height distribution, and no perfectly proportional friction law can be achieved. Green symbols identify target slopes satisfying all three criteria simultaneously, and thus constituting the validated \emph{reachable space} for proportional laws in our framework.

\begin{figure}[b!]
\centering
\includegraphics[width=0.6\textwidth]{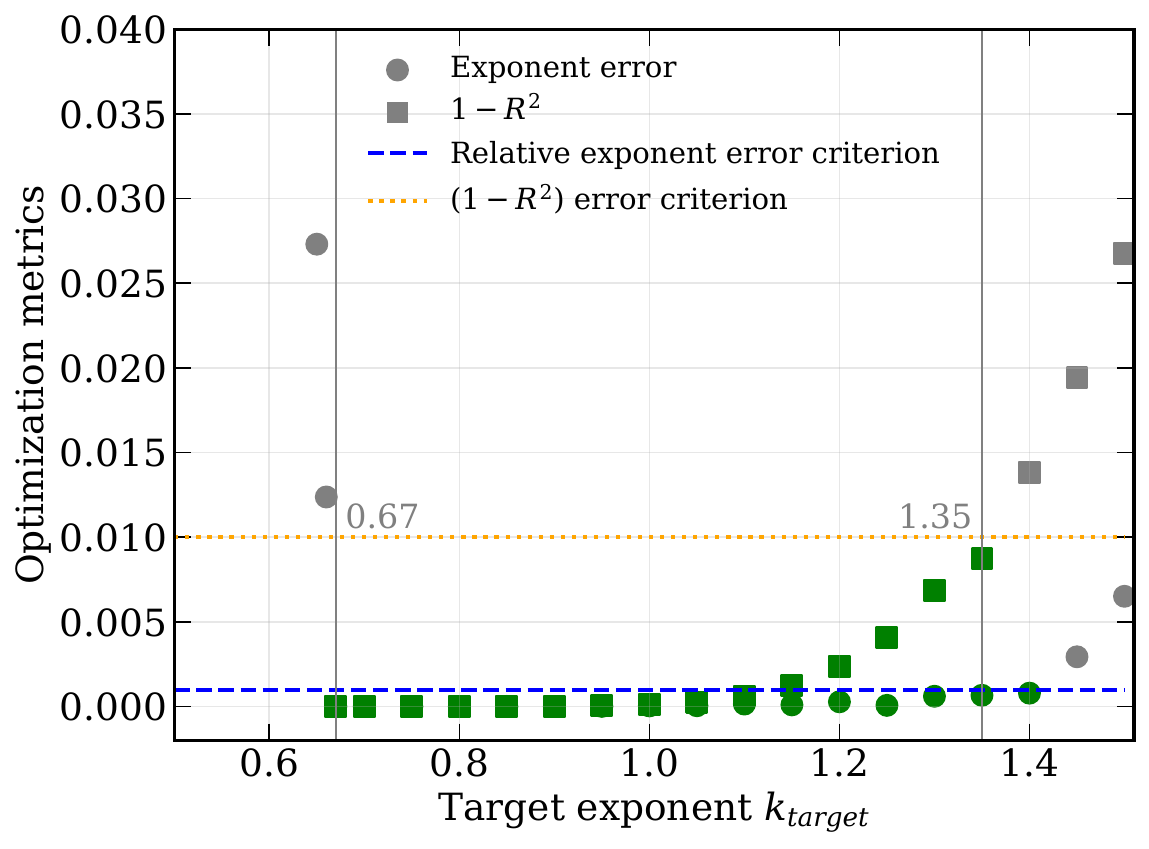}
\caption{Optimisation metrics related to the exponent $k$ and the global $R^2$ goodness of fit vs the target exponent $k_{\text{target}}$ for Case 2 (power friction laws). 
A relative exponent error less than 0.1\% (dashed blue line) is achieved for all $k_{\text{target}}$ between 0.67 and 1.40. For $k_{\text{target}}$ between 0.67 and 1.35, the global $R^2$ goodness of fit is also larger than 0.99 (dashed orange line). Green symbols represent admissible target exponent value according to each criterion.}
\label{fig-case_2_metrics}
\end{figure}

For power laws, Fig.~\ref{fig-case_2_metrics} identifies sufficiently good solutions based on both the relative exponent error $|k - k_\text{target}|/k_\text{target}$ and the global goodness-of-fit $R^2$. A relative exponent error below 0.1\% is achieved for all $k_\text{target} \in [0.67, 1.40]$, while $R^2 > 0.99$ is satisfied over  $[0.67, 1.35]$. The common interval $k_\text{target} \in [0.67, 1.35]$, where both criteria are satisfied simultaneously, thus corresponds to the \emph{reachable space} for power-law friction laws in our framework.

\section{Expanded design space enabled by Cases 1 and 2}

\begin{figure}[h!]
\centering
\includegraphics[width=0.7\textwidth]{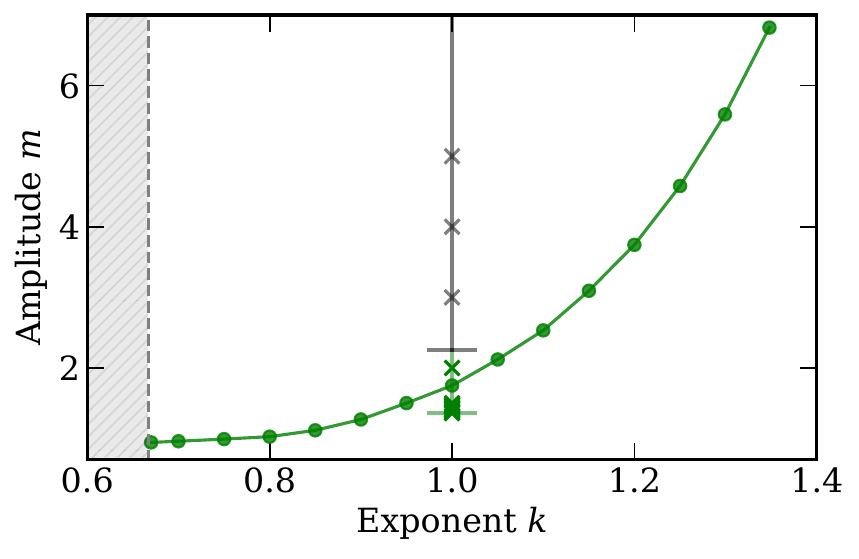}
\caption{Reachable space of friction parameters in the $(k,\,m)$ plane, where $\tilde{A} = m\,\tilde{P}^k$.
Green circles connected by a solid line show the best solutions unlocked by the GA for Case~2, which continuously span exponents from the Hertzian lower bound $k = 2/3$ (dashed vertical line; the gray-shaded region is physically inaccessible) up to $k \approx 1.35$.
Since a proportional law is a special case of a power law ($k = 1$), this reachable space also encompasses Case~1. The gray vertical segment at $k = 1$ indicates the range of amplitudes accessible with literature designs based on a truncated-exponential height distribution~\cite{aymard2024}, which are confined to proportional laws with amplitude $a \ge 2.25$. The present work on Case 1 (cross markers are the solutions actually obtained) expands the accessible design space toward lower slopes (green section of the vertical line), with a minimum achievable amplitude as low as $a \ge 1.36$.}
\label{fig:phase_diagram}
\end{figure}

Figure~\ref{fig:phase_diagram} offers a visual representation of the expansion of design possibilities enabled by our work on Cases 1 (proportional law) and 2 (power law). This representation is based on the common general expression of the dimensionless law in both cases: $\tilde{A} = m\,\tilde{P}^k$. Proportional laws correspond to $k$=1, with $m$ being the slope of the law (denoted as $a$ in Case 1). Figure~\ref{fig:phase_diagram} locates our new designs along the $(k,m)$ plane.

\textit{1D expansion: along the amplitude axis at $k$=1}\\
\noindent In~\cite{aymard2024}, quasi-linear friction laws were achieved using a truncated exponential height distribution. In our manuscript, we have shown that with such a distribution, proportional laws can be obtained only for slopes $m \ge 2.25$ (see the grey vertical segment in Fig.~\ref{fig:phase_diagram}).

Thanks to the TTE-based distributions identified for Case 1 through our GA-based design tool, we can now reach proportional laws with slopes down to $m$=1.36. In Fig.~\ref{fig:phase_diagram}, this 1-dimensional expansion of possibilities is represented by the green vertical segment that extends the grey one.

\textit{2D expansion: adding the exponent axis}\\
\noindent The satisfactory designs identified in Case 2 span a continuous range of exponents between 2/3 and 1.35, thus populating the $(k,m)$ plane with a non-vertical curve. In Fig.~\ref{fig:phase_diagram}, the region $k<$2/3 is shaded to show that no design could be identified there. While the case $k$=2/3 can arguably be considered an already existing design (Hertzian case), all the green points had never been found before.

\textit{Expansion beyond the $(k,m)$ plane}\\
\noindent The bilinear laws built within our Case 3 cannot be represented in Fig.~\ref{fig:phase_diagram}, but still represent a class of laws that could only be unlocked thanks to metainterfaces.

We emphasize that our GA-based design operates in the dimensionless space of $\tilde{A}$ and $\tilde{P}$, and not in the physical space of the dimensional $A$ and $P$. In this context, each point along the green curve in Fig.~\ref{fig:phase_diagram} demonstrates our capacity to create a dimensionless friction law with a specific (power law) shape. But virtually any desired dimensional amplitude can still be achieved for the same shape, without changing the identified height distribution, only by playing on the other system parameters. Those include both geometrical parameters $N$, $R$, $h_m$ and a material parameter, $E^*$, with the former being the most convenient to tune the amplitude of the law without changing the materials in contact.

This separation of roles is a key asset of the framework: the GA designs the functional form of the friction law in dimensionless space, while the practitioner navigates the dimensional parameter space (thanks to $E^*$, $h_\text{m}$, $R$ and $N$) to set the desired force scales. Together, these two complementary types of degrees of freedom provide access to a much broader design space than what is depicted in Fig.~\ref{fig:phase_diagram} alone.


\section{Tribological calibration of individual microcontacts}

\begin{figure}[h!]
\centering
\includegraphics[width=0.84\textwidth]{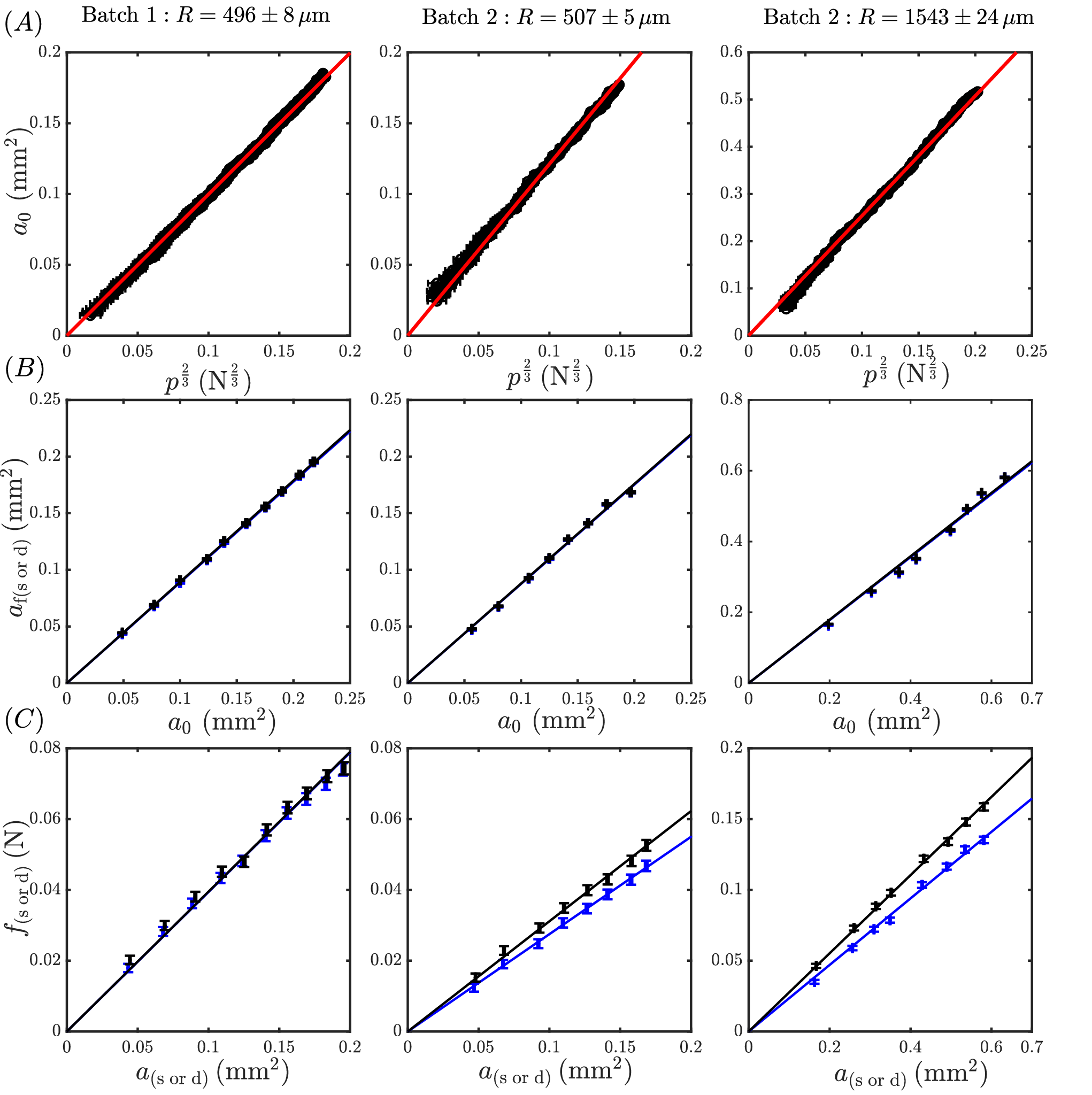}
\caption{Tribological calibration of single microcontacts, \REV{for both PDMS batches and both asperity radii. Left column: batch 1, nominal radius 500\,$\mu$m. Middle column: batch 2, nominal radius 500\,$\mu$m. Right column: batch 2, nominal radius 1500\,$\mu$m.} (A): $a_0$ vs $p^{2/3}$, for \REV{4 (left) or 3 (middle and right)} indentation experiments (symbols). Solid red lines: proportional fits enabling calibration of $E^*$ \REV{(see Tab.~\ref{Tab:Calib})}, using Hertz theory ($a_0=\pi \left(\frac{3Rp}{4E^*}\right)^{2/3}$) knowing $R$ \REV{(Tab.~\ref{Tab:Calib}).} (B): $a_{\text{f,s}}$ (black) and $a_{\text{f,d}}$ (blue) vs $a_0$ (symbols). Solid lines: proportional fits enabling calibration of $B_{\text{s}}$ and $B_{\text{d}}$ \REV{(Tab.~\ref{Tab:Calib}).} (C): Symbols: $f_{\text{s}}$  vs $a_{\text{f,s}}$ (black) and $f_{\text{d}}$  vs $a_{\text{f,d}}$ (blue). Solid lines: proportional fits, showing the existence of asperity-level friction stresses, $\sigma_\text{s}$ and $\sigma_\text{d}$.}
\label{fig-calib}
\end{figure}

\begin{table*}[h!]
\centering
\caption{\REV{Calibrated curvature radius, elastic modulus and area reduction ratii used to calculate the dimensionless quantities in all experimental figures. For Fig.~4B, the values of $E^*$, $B_\text{s}$ and $B_\text{d}$ are averages over the two nominal radii, which are found to match within the error bars. For each value, the number of underlying measurements is indicated in brackets. $\pm$ denotes 95\% confidence interval.}}
\resizebox{\columnwidth}{!}{%
\begin{tabular}{l|c|c|c}
 & Figs.~2B and 3B ($k_{\text{target}}$=0.85) & Figs.~3B ($k_{\text{target}}$=0.70 and 1.20) &  Fig.~4B\\
\hline
 & Batch 1, Radius 1 & Batch 2, Radius 1 & Batch 2, Radii 1 \& 2\\
\hline
$R$ ($\mu$m) & 496 $\pm$8 (5) & 507$\pm$5 (5) & 507$\pm$5 (5) \& 1543$\pm$24 (5) \\
$E^*$ (MPa) & 2.08$\pm$0.08 (4) & 1.59$\pm$0.02 (3) & 1.59$\pm$0.01 (3+3)\\
$B_\text{s}$ & 0.894$\pm$0.003 (1) & 0.879$\pm$0.017 (1) & 0.883$\pm$0.015 (1+1)\\
$B_\text{d}$ & 0.889$\pm$0.004 (1) & 0.876$\pm$0.017 (1) & 0.879$\pm$0.015 (1+1)\\
\hline
\end{tabular}\label{Tab:Calib}
}
\end{table*}

\begin{table*}[h!]
\centering
\caption{\REV{Values of the friction stresses, $\sigma_{\text{s}}$ and $\sigma_{\text{d}}$, used to calculate the dimensionless friction forces, $\tilde{F}_{\text{s}}$ and $\tilde{F}_{\text{d}}$ respectively, in all experimental figures (1 measurement). $\pm$ denotes 95\% confidence interval.}}
\resizebox{0.5\columnwidth}{!}{%
\begin{tabular}{l|cc}
Experimental curve & $\sigma_{\text{s}}$ (MPa) & $\sigma_{\text{d}}$ (MPa)\\
\hline
Fig.~2B & 0.355$\pm$0.019 & 0.365$\pm$0.016 \\
Fig.~3B, $k_{\text{target}}$=0.70 & 0.304$\pm$0.005 & 0.269$\pm$0.015 \\
Fig.~3B, $k_{\text{target}}$=0.85 & 0.330$\pm$0.011 & 0.321$\pm$0.011 \\
Fig.~3B, $k_{\text{target}}$=1.20 & 0.305$\pm$0.005 & 0.307$\pm$0.007 \\
Fig.~4B & 0.280$\pm$0.006 & 0.304$\pm$0.013 \\

\hline
\end{tabular}\label{Tab:Calibsigma}
}
\end{table*}


\section{Precautions to avoid elastic interactions between microcontacts}

The friction model of Eqs.~1-2 assumes that microcontacts are elastically independent. To investigate the conditions in which this assumption is valid, the authors of~\cite{zeka2026} have performed a finite element study of metainterfaces similar to that used in the present experiments. They could conclude on recommendations about the conditions in which no measurable effect of elastic interactions is expected. We have been careful to use experimental parameters that match those recommendations. Beyond the large pitch of the square lattice (1.5\,mm) that ensures a sufficient distance between microcontacts, we have avoided clusters of high asperities and used a sufficiently large size of the asperity-bearing parallelepipedic bulk: a thickness of 7.2\,mm and a width 8\,mm larger than the side length of the central square lattice.

\section{Lists of prescribed heights and radii of experimental metainterfaces}

\begin{table*}[h!]
\centering
\caption{List of prescribed heights for the  metainterface with a proportional friction law with target slope $a_{\text{target}}$=1.39 (Fig.~2B). For manufacturing, a common offset $h_{\text{0}}$=50$\mu$m is added to all values in the table.}
\resizebox{\columnwidth}{!}{%
\begin{tabular}{l|cccccccccccccccccccccc}
Asperity & 1 & 2 & 3 & 4 & 5 & 6 & 7 & 8 & 9 & 10 & 11 & 12 & 13 &  14 & 15 & 16\\
\hline
$h_i$ ($\mu$m) & 0.5 & 1.1 & 10.3 & 33.9 & 23.6 & 16.7 & 114.3 & 29.6 & 42.6 & 57.9 & 37.1 & 44.7 & 28.3 & 14.2 & 69.7 & 3.3\\
\hline
Asperity & 17 & 18 & 19 & 20 & 21 & 22 & 23 & 24 & 25 & 26 & 27 & 28 & 29 & 30 & 31 & 32\\
\hline
$h_i$ ($\mu$m) & 61.3 & 65.2 & 18.5 & 5.0 & 74.8 & 51.9 & 12.6 & 31.0 & 8.9 & 9.6 & 11.1 & 5.7 & 99.0 & 3.9 & 54.8 & 2.1 \\
\hline
Asperity & 33 & 34 & 35 & 36 & 37 & 38 & 39 & 40 & 41 & 42 & 43 & 44 & 45 & 46 & 47 & 48\\
\hline
$h_i$ ($\mu$m) & 117.7 & 49.3 & 20.5 & 32.4 & 15.8 & 22.5 & 19.5 & 119.1 & 1.6 & 38.8 & 27.1 & 35.5 & 2.7 & 6.3 & 40.7 & 4.4 \\
\hline
Asperity & 49 & 50 & 51 & 52 & 53 & 54 & 55 & 56 & 57 & 58 & 59 & 60 & 61 & 62 & 63 & 64 \\
\hline
$h_i$ ($\mu$m) & 11.8 & 119.6 & 17.6 & 81.0 & 25.9 & 8.2 & 24.7 & 15.0 & 118.5 & 7.6 & 46.9 & 21.5 & 120.0 & 88.7 & 13.4 & 6.9 \\
\hline
\end{tabular}\label{Tab:hprop}
}
\end{table*}

\begin{table*}[h!]
\centering
\caption{\REV{List of prescribed heights for the  metainterface with a power friction law with a target exponent $k_{\text{target}}$=0.70 (Fig.~3B). For manufacturing, a common offset $h_{\text{0}}$=50$\mu$m is added to all values in the table.}}
\resizebox{\columnwidth}{!}{%
\begin{tabular}{l|cccccccccccccccccccccc}
Asperity & 1 & 2 & 3 & 4 & 5 & 6 & 7 & 8 & 9 & 10 & 11 & 12 & 13 &  14 & 15 & 16\\
\hline
$h_i$ ($\mu$m) & 82.3 & 80.4 & 89.1 & 56.7 & 0.5 & 87.1 & 15.4 & 111.1 & 105.4 & 117.6 & 119.5 & 118.6 & 117.9 & 118.0 & 118.5 & 112.8\\
\hline
Asperity & 17 & 18 & 19 & 20 & 21 & 22 & 23 & 24 & 25 & 26 & 27 & 28 & 29 & 30 & 31 & 32\\
\hline
$h_i$ ($\mu$m) & 0.1 & 117.5 & 117.7 & 119.3 & 114.7 & 119.7 & 118.1 & 67.5 & 62.2 & 119.3 & 113.5 & 118.0 & 120.0 & 116.4 & 119.7 & 104.4 \\
\hline
Asperity & 33 & 34 & 35 & 36 & 37 & 38 & 39 & 40 & 41 & 42 & 43 & 44 & 45 & 46 & 47 & 48\\
\hline
$h_i$ ($\mu$m) & 35.6 & 118.4 & 119.7 & 114.5 & 118.7 & 119.9 & 119.6 & 111.3 & 8.8 & 119.9 & 116.5 & 119.7 & 119.8 & 115.8 & 118.1 & 112.6 \\
\hline
Asperity & 49 & 50 & 51 & 52 & 53 & 54 & 55 & 56 & 57 & 58 & 59 & 60 & 61 & 62 & 63 & 64 \\
\hline
$h_i$ ($\mu$m) & 40.3 & 117.2 & 119.6 & 119.0 & 118.2 & 118.9 & 119.4 & 7.7 & 78.3 & 93.7 & 103.5 & 112.3 & 99.7 & 96.9 & 51.9 & 3.0 \\
\hline
\end{tabular}\label{Tab:hpow07}
}
\end{table*}

\begin{table*}[h!]
\centering
\caption{List of prescribed heights for the  metainterface with a power friction law with a target exponent $k_{\text{target}}$=0.85 (Fig.~3B). For manufacturing, a common offset $h_{\text{0}}$=50$\mu$m is added to all values in the table.}
\resizebox{\columnwidth}{!}{%
\begin{tabular}{l|cccccccccccccccccccccc}
Asperity & 1 & 2 & 3 & 4 & 5 & 6 & 7 & 8 & 9 & 10 & 11 & 12 & 13 &  14 & 15 & 16\\
\hline
$h_i$ ($\mu$m) & 9.9 & 21.0 & 48.7 & 98.9 & 9.4 & 28.1 & 74.0 & 91.9 & 116.2 & 25.6 & 47.5 & 86.7 & 39.4 & 31.4 & 5.1 & 88.7\\
\hline
Asperity & 17 & 18 & 19 & 20 & 21 & 22 & 23 & 24 & 25 & 26 & 27 & 28 & 29 & 30 & 31 & 32\\
\hline
$h_i$ ($\mu$m) & 95.5 & 57.6 & 52.3 & 39.6 & 103.4 & 86.9 & 67.1 & 24.6 & 34.5 & 3.8 & 67.8 & 51.4 & 53.7 & 94.5 & 113.1 & 9.6 \\
\hline
Asperity & 33 & 34 & 35 & 36 & 37 & 38 & 39 & 40 & 41 & 42 & 43 & 44 & 45 & 46 & 47 & 48\\
\hline
$h_i$ ($\mu$m) & 42.9 & 16.6 & 62.8 & 4.4 & 53.1 & 55.3 & 39.5 & 22.8 & 41.5 & 81.2 & 25.1 & 7.3 & 95.4 & 80.5 & 119.4 & 45.1 \\
\hline
Asperity & 49 & 50 & 51 & 52 & 53 & 54 & 55 & 56 & 57 & 58 & 59 & 60 & 61 & 62 & 63 & 64 \\
\hline
$h_i$ ($\mu$m) & 17.6 & 51.6 & 59.9 & 31.2 & 64.9 & 116.0 & 22.4 & 41.9 & 1.2 & 44.2 & 18.2 & 70.4 & 44.3 & 24.9 & 26.9 & 18.6 \\
\hline
\end{tabular}\label{Tab:hpow085}
}
\end{table*}

\begin{table*}[h!]
\centering
\caption{\REV{List of prescribed heights for the  metainterface with a power friction law with a target exponent $k_{\text{target}}$=1.20 (Fig.~3B). For manufacturing, a common offset $h_{\text{0}}$=50$\mu$m is added to all values in the table.}}
\resizebox{\columnwidth}{!}{%
\begin{tabular}{l|cccccccccccccccccccccc}
Asperity & 1 & 2 & 3 & 4 & 5 & 6 & 7 & 8 & 9 & 10 & 11 & 12 & 13 &  14 & 15 & 16\\
\hline
$h_i$ ($\mu$m) & 6.6 & 3.7 & 2.7 & 7.1 & 7.7 & 2.3 & 4.7 & 4.8 & 6.5 & 9.4 & 17.5 & 48.3 & 11.5 & 12.6 & 12.4 & 4.1\\
\hline
Asperity & 17 & 18 & 19 & 20 & 21 & 22 & 23 & 24 & 25 & 26 & 27 & 28 & 29 & 30 & 31 & 32\\
\hline
$h_i$ ($\mu$m) & 5.2 & 22.0 & 9.4 & 14.5 & 17.3 & 29.6 & 17.2 & 6.1 & 8.3 & 12.7 & 32.3 & 17.7 & 119.3 & 11.4 & 41.2 & 8.2 \\
\hline
Asperity & 33 & 34 & 35 & 36 & 37 & 38 & 39 & 40 & 41 & 42 & 43 & 44 & 45 & 46 & 47 & 48\\
\hline
$h_i$ ($\mu$m) & 8.8 & 25.3 & 11.2 & 11.2 & 11.6 & 16.2 & 9.8 & 6.5 & 8.8 & 9.3 & 45.5 & 9.9 & 26.9 & 11.6 & 35.5 & 8.6 \\
\hline
Asperity & 49 & 50 & 51 & 52 & 53 & 54 & 55 & 56 & 57 & 58 & 59 & 60 & 61 & 62 & 63 & 64 \\
\hline
$h_i$ ($\mu$m) & 3.7 & 18.9 & 23.0 & 21.2 & 10.8 & 26.9 & 9.4 & 5.3 & 4.4 & 6.1 & 6.2 & 4.7 & 7.1 & 6.8 & 5.7 & 4.6 \\
\hline
\end{tabular}\label{Tab:hpow1p2}
}
\end{table*}

\begin{table*}[h!]
\centering
\caption{\REV{List of prescribed heights and radii for the metainterface with a bilinear friction law (Fig.~4B). Radii are expressed in units (1 or 3) of the smallest nominal radius, $R$=500\,$\mu$m. For manufacturing, a common offset $h_{\text{0}}$=50$\mu$m is added to all height values in the table.}}
\resizebox{\columnwidth}{!}{%
\begin{tabular}{l|cccccccccccccccccccccc}
Asperity & 1 & 2 & 3 & 4 & 5 & 6 & 7 & 8 & 9 & 10 & 11 & 12 & 13 &  14 & 15 & 16\\
\hline
$h_i$ ($\mu$m) & 5.3 & 3.8 & 6.0 & 0.9 & 0.0 & 5.7 & 0.2 & 10.0 & 8.8 & 84.9 & 94.3 & 19.1 & 75.2 & 81.2 & 81.3 & 13.7\\
$R_i/R$ & 1 & 1 & 3 & 1 & 1 & 1 & 3 & 3 & 3 & 1 & 3 & 3 & 1 & 1 & 1 & 3\\
\hline
Asperity & 17 & 18 & 19 & 20 & 21 & 22 & 23 & 24 & 25 & 26 & 27 & 28 & 29 & 30 & 31 & 32\\
\hline
$h_i$ ($\mu$m) & 0.0 & 80.7 & 36.3 & 28.4 & 118.5 & 117.3 & 78.2 & 2.9 & 0.9 & 83.7 & 118.9 & 27.2 & 20.5 & 29.8 & 21.3 & 6.9 \\
$R_i/R$ & 3 & 1 & 1 & 1 & 1 & 1 & 1 & 1 & 3 & 1 & 1 & 1 & 1 & 1 & 1 & 1\\
\hline
Asperity & 33 & 34 & 35 & 36 & 37 & 38 & 39 & 40 & 41 & 42 & 43 & 44 & 45 & 46 & 47 & 48\\
\hline
$h_i$ ($\mu$m) & 0.5 & 76.4 & 31.6 & 14.6 & 22.3 & 22.5 & 119.6 & 11.5 & 0.1 & 77.5 & 17.2 & 89.9 & 119.8 & 93.1 & 21.7 & 13.0 \\
$R_i/R$ & 3 & 1 & 3 & 3 & 3 & 1 & 1 & 3 & 3 & 1 & 1 & 1 & 3 & 1 & 3 & 3\\
\hline
Asperity & 49 & 50 & 51 & 52 & 53 & 54 & 55 & 56 & 57 & 58 & 59 & 60 & 61 & 62 & 63 & 64 \\
\hline
$h_i$ ($\mu$m) & 0.6 & 76.6 & 92.5 & 15.2 & 30.7 & 14.4 & 70.7 & 0.1 & 3.3 & 6.4 & 6.7 & 12.8 & 6.7 & 6.4 & 0.6 & 0.0 \\
$R_i/R$ & 3 & 1 & 1 & 1 & 1 & 3 & 1 & 3 & 3 & 1 & 3 & 1 & 1 & 1 & 1 & 1\\
\hline
\end{tabular}\label{Tab:hbilin}
}
\end{table*}

\newpage

\section{Sensitivity of friction laws to manufacturing errors}

The fabrication errors are estimated by comparing the asperity heights measured on the sample used in Fig.~2B (proportional law) and the prescribed heights from the design phase. This difference includes all errors, from mould manufacturing to PDMS replication. The heights are obtained from a spherical fit of the 3D topography of each of the 64 asperities. The topography is measured with an interferometric profilometer (Bruker Contour GT-K1), after deposition of a thin film (about 5\,nm) of a mixture of gold (60\% mass) and palladium (40\% mass) at the surface of the PDMS sample, to increase its reflectivity. The fit has the summit height as the only fitting parameter, while the asperity radius $R$ is set to its previously calibrated value, 496\,$\mu$m. The standard deviation of the 64 height differences is found less than 2.1\,$\mu$m.

\begin{figure}[h!]
     \centering
     \includegraphics[width=0.6\linewidth]{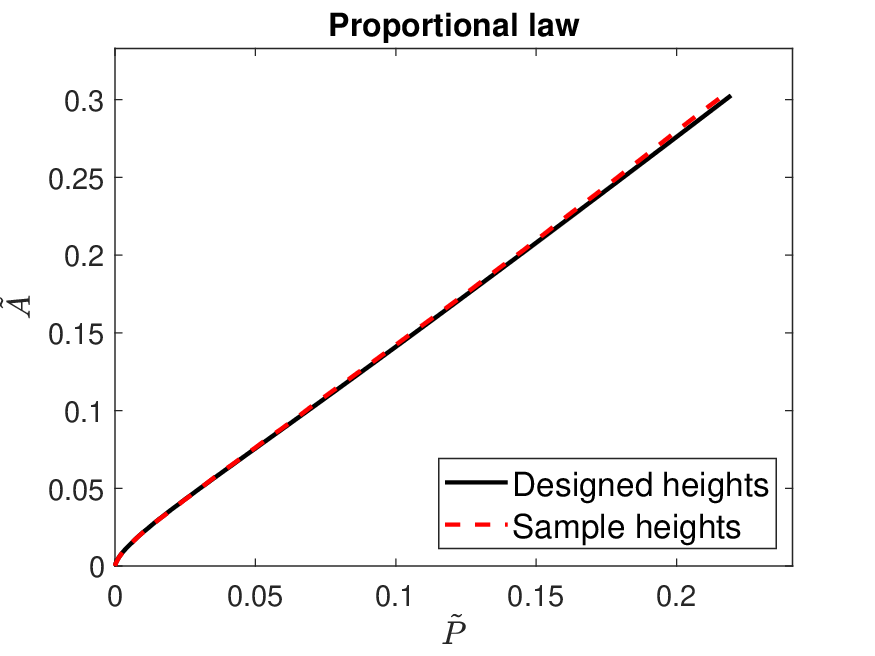}
        \caption{Predicted friction law for two height lists in the case of the proportional law. Black solid line: using the as-designed heights of Tab.~\ref{Tab:hprop}. Red dashed line: using the heights measured on the experimental sample.}
\label{fig:ManufErrorFrictionLaw}
\end{figure}

To visualize the impact of fabrication errors on the friction law, we compare in Fig.~\ref{fig:ManufErrorFrictionLaw} the proportional friction laws predicted by the contact model for two height lists: (i) that from the design phase, as given in Tab.~\ref{Tab:hprop} and (ii) that from the profilometry measurements. As can be seen, the difference is only marginal: the root mean square of the difference is just 1.0\% of the root mean square of the designed friction law.

To explore more systematically the impact of manufacturing errors on the resulting friction law, we perform a Monte-Carlo analysis. We start with the as-designed lists of heights for both the proportional law shown in Fig.~2B of the article (Tab.~\ref{Tab:hprop}) and the power law with exponent 0.85 shown in Fig.~3B of the article (Tab.~\ref{Tab:hpow085}). We then add a random centered gaussian noise, $\Delta h$, on all heights and use the contact model to predict the modified friction law. We finally fit the friction law and extract relevant quantifier of the shape. This procedure is repeated for different values of the standard deviation of the gaussian noise, and for 10000 different realizations of the noise for each standard deviation. 

\begin{figure}[h!]
     \centering
     \includegraphics[width=0.49\linewidth]{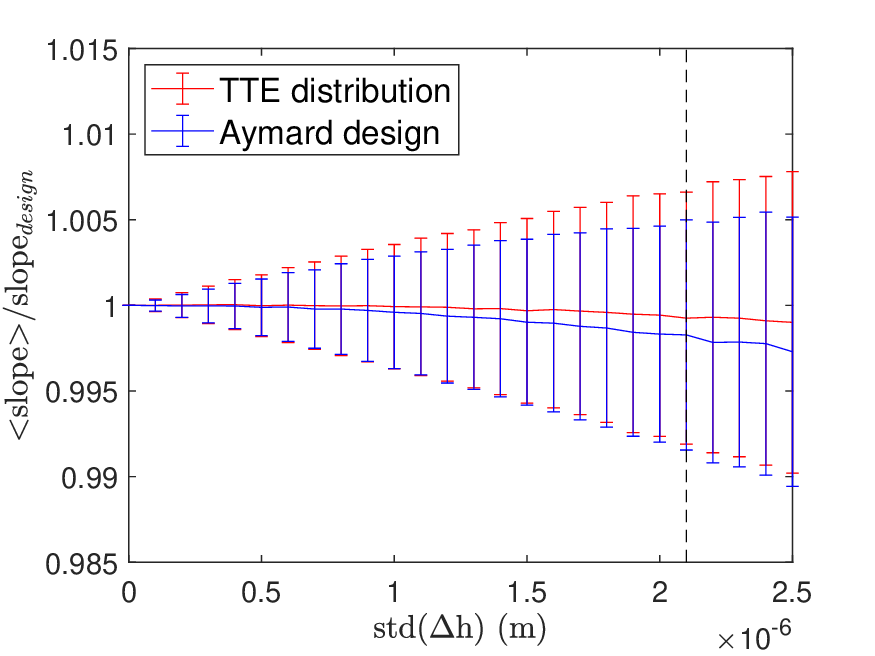}
     \includegraphics[width=0.49\linewidth]{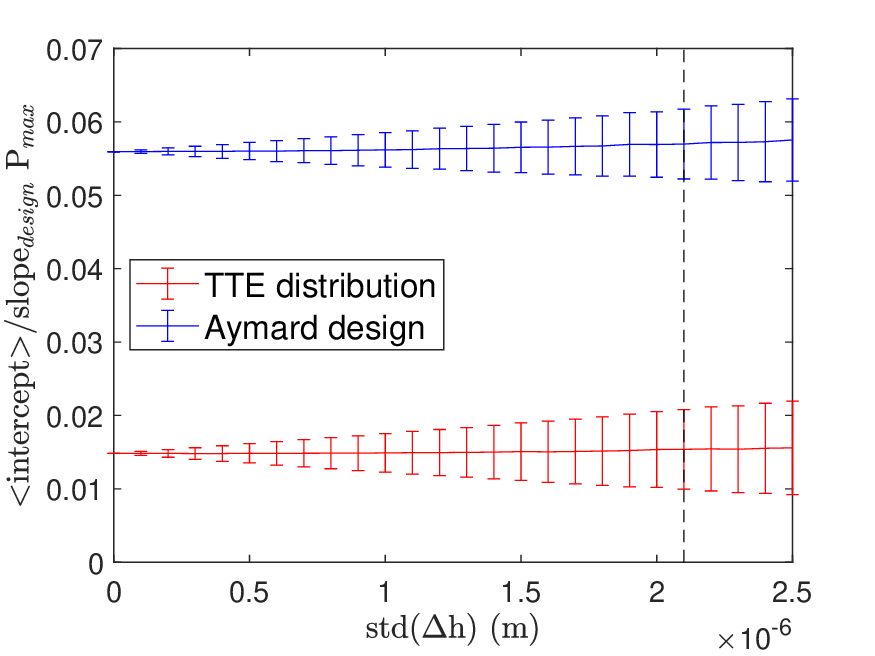}
        \caption{Evolution of the mean asymptotic slope (left) and intercept (right) of the predicted proportional friction law, as functions of the standard deviation of the noise added to the asperity heights. Red: the reference heights are those obtained using the triangular-based truncated exponential (TTE) and underlying Fig.~2B of the article (Tab.~\ref{Tab:hprop}). Blue: the reference heights are those obtained using a simple truncated exponential and underlying the red line in Fig.~4 of~\cite{aymard2024}. For each point, 10000 different realizations of the noise are considered. Error bars represent the standard deviation of the fitted parameters. The vertical dashed line represents the manufacturing precision in the experimental realization of the TTE-based height distribution.}
\label{fig:MCProp}
\end{figure}

For the proportional law, the fit is a linear fit over the last 20\% of the points. Figure~\ref{fig:MCProp} represents, in red, the evolution of the mean and standard deviation of both the fitted slope and intercept, as a function of the standard deviation of the noise. For comparison, the same quantities are shown, in blue, for the quasi-linear design of~\cite{aymard2024} (see red line in Fig.~4 therein). For both designs, the mean of the slope varies only marginally, and the standard deviation steadily increases as the amplitude of the manufacturing error increases. For the experimental error of 2.1\,$\mu$m (dashed vertical line), the standard deviation of the perturbed slope is less than 0.85\% of the as-designed slope (Fig.~\ref{fig:MCProp}, left). The mean intercept is also essentially insensitive to the manufacturing errors, while its standard deviation steadily increases. For our experimental level of error, the sum of the mean and standard deviation remains smaller than 2.1\% of the product of the as-designed slope and the maximum explored normal force (red data in Fig.~\ref{fig:MCProp}, right). Importantly, this value remains always about three times smaller than for the literature design (blue data in Fig.~\ref{fig:MCProp}, right). This shows that, even when considering manufacturing errors, the proposed new design significantly outperforms the existing designs.

\begin{figure}[h!]
     \centering
     \includegraphics[width=0.49\linewidth]{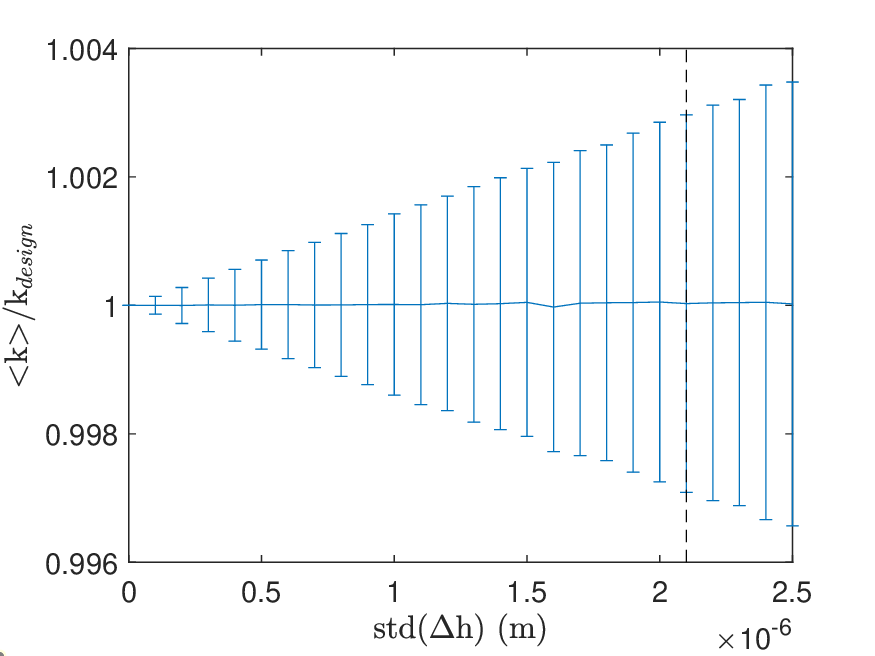}
     \includegraphics[width=0.49\linewidth]{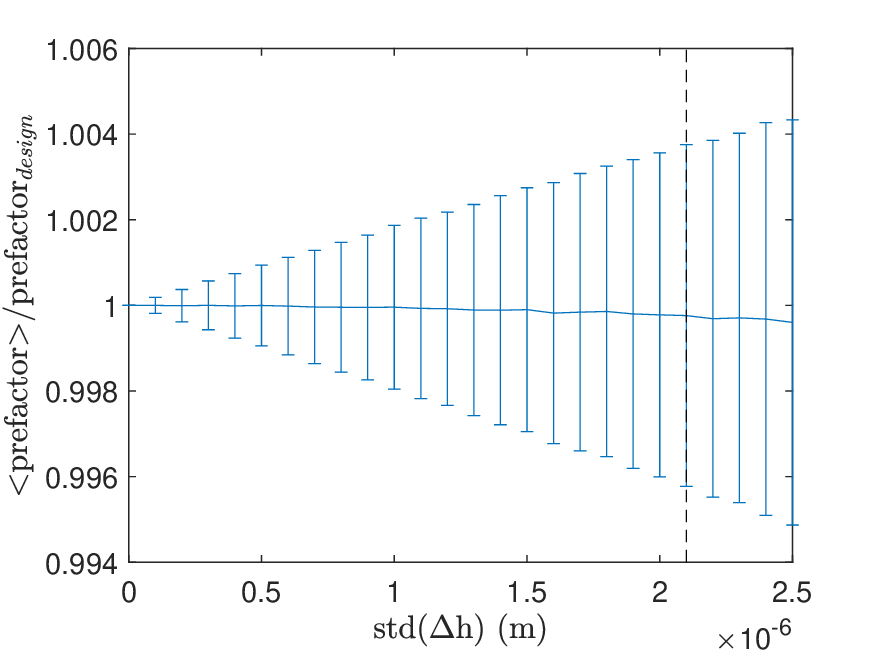}
        \caption{Evolution of the mean exponent $k$ (left) and prefactor (right) of the predicted power friction law, as functions of the standard deviation of the noise added to the asperity heights. The reference heights are those underlying Fig.~3B of the article (Tab.~\ref{Tab:hpow085}). For each point, 10000 different realizations of the noise are considered. Error bars represent the standard deviation of the fitted parameters. The vertical dashed line represents the manufacturing precision in the experimental realization.}
\label{fig:MCPower}
\end{figure}

For the power law with exponent 0.85, the fit is a power fit over all data points, with both the exponent and prefactor being fitting parameters. Figure~\ref{fig:MCPower} represents, in a way analogous to Fig.~\ref{fig:MCProp}, the evolution of the mean and standard deviation of both fitting parameters, as a function of the standard deviation of the noise. It appears that both the mean prefactor and mean exponent deviate negligibly from their as-designed values. For the experimental level of error (vertical dashed line), the standard deviation is less that 0.43\% for both parameters. This indicates that the corresponding power friction law shown in Fig.~3B of the article is particularly immune to manufacturing errors.

\section{Absence of evolution of the interface during our measurements of the behaviour laws}

\begin{figure}[h!]
     \centering
     \includegraphics[width=\linewidth]{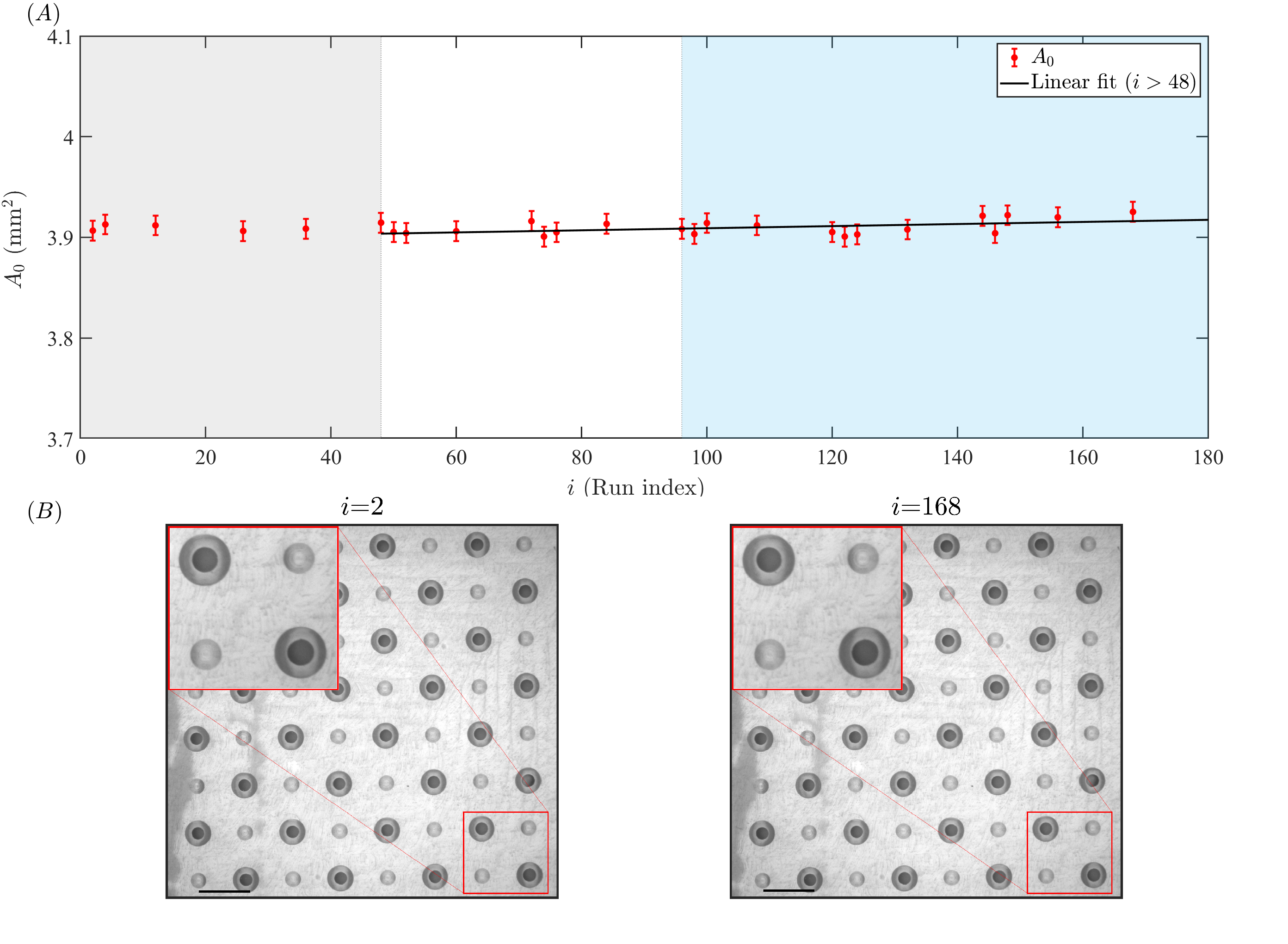}
             \caption{A: Evolution of the initial contact area, $A_0$, across 168 runs with identical sliding distance (1\,mm), sliding speed (0.1\,mm/s) and normal force (1.4\,N), on a specific sample (see B). Red points are measurements performed on the runs during which images have been acquired. Grey region: 48 first runs corresponding to run-in. White: next 48 runs during which tribological measurements are typically made. Blue: additional runs to assess potential wear-induced trends. Black line: linear fit over the white and blue regions. B: photographs of the interfaces at run indices 2 and 168. Red squares are zooms on 4 asperities. Scale bar: 1.5\,mm. No noticeable difference can be observed between the first and last runs.}
\label{fig:Wear}
\end{figure}

The measurement of each experimental curve in panels B of Figs.~2--4 involves a few tens of successive runs (one for each value of $\tilde{P}$). To assess whether wear or durabilty issues could impact those measured behaviour laws, we perform a dedicated experiment with a specific sample. This sample features 32 asperities of height 200\,$\mu$m, and 32 asperities with a lower height of 60\,$\mu$m, organised as a chessboard (high and small asperities alternate on the square lattice, see panel B of Fig.~\ref{fig:Wear}). We then perform 168 sliding runs, all with the same sliding distance and the same sliding speed as in our main experiments. The normal force is chosen such that only the 32 tallest asperities are in contact (indentation about 120\,$\mu$m, similar to the maximum indentation used when measuring friction laws).

The first 48 runs (grey region in Fig.~\ref{fig:Wear}A) correspond to the typical runs over which we run-in any newly prepared metainterface. Runs between 49 and 96 (white region) correspond to the typical runs necessary to measure the behaviours laws of each of the metainterfaces in panels B of Figs.~2 to 4. The rest of the runs (blue region) are to evaluate the potential longer-term evolution of the interface.

For some of the runs, contact images and/or tangential force measurements are performed, enabling evaluation of the stability of the areas and friction forces as the index of the runs increases. As an example, Fig.~\ref{fig:Wear}A shows the evolution of the initial contact area, $A_\text{0}$ as a function of the run index. As can be seen, no significant evolution can be seen over the whole experiment. To be more quantitative, we perform a linear fit of the data excluding the run-in phase (black line). This fit is then used to estimate the expected relative evolution of $A_\text{0}$ over the white region, i.e. that in which we typically measure the behaviour laws. According to this fit, $A_\text{0}$ varies by less than 0.2\% between runs 49 and 96, indicating that this evolution is unlikely to have any impact on the measured laws. Making identical estimates on the other interfacial measurements shows that negligible evolutions occur over the typical 48 runs necessary to build their respective behaviour laws. The expected relative evolutions are found less than 0.002, 0.09, 1.0 and 1.6\% for $A_\text{s}$, $A_\text{d}$, $F_\text{s}$ and $F_\text{d}$, respectively.

The absence of evolution of the interface along the 168 runs is illustrated in Fig.~\ref{fig:Wear}B, where we compare interface images at the beginning (run 2) and the end (run 168) of the sequence of runs. No noticeable change can be observed.

To be more quantitative, after the runs, we perform additional interferometric profilometry measurements on all asperities of the sample. Note that the 32 smallest asperities have been placed to serve as reference, un-worn asperities because for the normal force used, they have never been involved in the contacts. 3D profiles of each of the 64 asperities are fitted with a sphere with the calibrated radius $R$=507$\mu$m and the fitted summit height is extracted. We find that the heights of the 32 small (resp. high) asperities are 62$\pm$2$\mu$m (resp. 198$\pm$4$\mu$m). These values match the prescribed ones (200 and 60\,$\mu$m) to within the error bars, meaning that no geometrical change due to potential wear of the asperities could be detected.

Overall, the above results indicate that wear has no impact on the experimental results shown in panels B of Figs.~2 to 4.








\bibliography{GA_metainterface}

\end{document}